\documentclass[showpacs, pra,twocolumn,preprintnumbers ,amsmath, amssymb, superscriptaddress, aps]{revtex4-2}
\usepackage{color}
\usepackage{amsmath,amssymb}
\usepackage{pifont}
\usepackage{amssymb}  
\usepackage{bbold}
\usepackage{float}
\usepackage{placeins}
\usepackage{subfloat}
\usepackage{adjustbox}

\usepackage[caption=false]{subfig}
\usepackage{tikz}
\usepackage{makecell}
\usepackage{subfig}
\usepackage{pifont}  
\usepackage{graphicx}

\graphicspath{{Figuress/}}
\usepackage{dcolumn}  
\usepackage{bm}       
\usepackage{multirow} 
\usepackage[colorlinks]{hyperref}
\usepackage{mathtools}
\usepackage{appendix}

\newcommand{\ket}[1]{\left|{#1}\right\rangle}

\newcommand{\tr}{\mathrm{tr}}

\def \be{\begin{align}}
	\def \ee{\end{align}}
\def \bea{\begin{eqnarray}}
	\def \eea{\end{eqnarray}}

\begin{document}

	\title{
		Quantum correlations and coherence in a two-qubit anisotropic $XY$\\ under magnetic field}

\author{Ahmed Jellal}
\email{a.jellal@ucd.ac.ma}
\affiliation{Laboratory of Theoretical Physics, Faculty of Sciences, Choua\"ib Doukkali University, PO Box 20, 24000 El Jadida, Morocco}

\author{Pablo Díaz}
\affiliation{Departamento de Ciencias F\'{i}sicas, Universidad de La Frontera, Casilla 54-D, Temuco 4811230, Chile}

\author{David Laroze}
\affiliation{Instituto de Alta Investigación, Universidad de Tarapacá, Casilla 7D, Arica, Chile}

			\begin{abstract}
				
We study thermal quantum correlations and coherence in  Heisenberg $XY$ model with anisotropic interactions under a uniform magnetic field \( B \). Using concurrence \( C \), local quantum uncertainty (LQU),  Bell-Clauser-Horne-Shimony-Holt (CHSH) nonlocality \( \mathbb{B} \), and coherence \( C_l \) as quantifiers, we analyze how magnetic anisotropy \( \delta_m \), coupling anisotropy \( \delta_c \), Dzyaloshinskii-Moriya (DM) interaction \( D \), temperature \( T \), and magnetic field \( B \) modulate quantum resources. At low temperatures  and relevant magnetic fields, the entanglement is maximized, but exhibits sudden death for \( \delta_m = 0 \), which turns into a smooth decay as \( \delta_m \) increases, highlighting its stabilizing role. LQU shows that stronger anisotropy suppresses quantum correlations, while \( \mathbb{B} \) induces a non-monotonic response peaking at a critical field \( B_c \). Bell-CHSH nonlocality violations (\( \mathbb{B} > 2 \)) persist below \( B_c \), but thermal noise (\( T \geq 1 \)) suppresses them. Coherence \( C_l \) is most robust to thermal fluctuations, especially for high \( \delta_m \), which also dampens abrupt quantum phase transitions. The DM interaction is essential for entanglement generation, with \( D \) and anisotropy synergistically enhancing correlation resilience. We identify a hierarchy of thermal degradation: nonlocality (\( \mathbb{B} \)) vanishes first, followed by entanglement (\( C \)), then general quantum correlations (LQU), while coherence \( C_l \) persists the longest. These results demonstrate tunable control of quantum resources via anisotropy and external parameters, providing insights for the design of robust spin-based quantum technologies.

				
			\end{abstract}

		\maketitle

	\section{Introduction}

	In quantum information theory, correlations can be either local or non-local. In the former category, correlations can be quantified using local quantum uncertainty (LQU) \cite{LQU0, Adesso} or local quantum Fisher information (LQFI) \cite{LQFI}. These two quantifiers characterize the sensitivity of a system to local perturbations. The second category is described by entanglement measures, such as concurrence \cite{Woo}, negativity \cite{Niga}, and  violations of Bell-Clauser-Horne-Shimony-Holt (CHSH) inequalities \cite{Bell, Bell-CHSH}. These measures demonstrate the distinctly non-classical nature of quantum systems.
	Quantum coherence is another important quantum resource. It is quantified by measures such as the $C_l$ norm \cite{Coherence}. Quantum correlations \cite{correlation} and coherence \cite{Coherence0} serve as critical resources for various quantum information processing tasks. Examples include quantum cryptography \cite{app1}, superdense coding \cite{app2}, quantum key distribution \cite{app3}, and quantum teleportation \cite{app4}. These applications exploit the unique advantages of quantum systems to achieve performance beyond classical limits.
	An important problem in the field is {how to maintain and deepen quantum correlations  in order to enable reliable communication between quantum systems}. {Recently, several promising proposals have been investigated in the context of preserving and strengthening these correlations. Among these are}
	the Dzyaloshinsky-Moriya (DM) interaction \cite{QC00}, and anisotropic coupling effects \cite{Anisotropy}. These achievements not only further the understanding of the fundamental properties of quantum systems but are also stepping stones toward the developing  strong quantum communication protocols. {Studying local}, and nonlocal quantum correlations {helps bridge the gap between} theory and real-world technological possibilities in quantum information science.


	Two-qubit systems play a central role in quantum information science because they represent the minimal units capable of generating and manipulating entanglement, which is an essential resource for quantum computation, teleportation, and secure communication \cite{NielsenChuang2011,Horodecki2009}. While many-body techniques offer powerful numerical capabilities, such systems often obscure physical mechanisms due to their complexity. By contrast, two-qubit models, such as the anisotropic $XY$ Hamiltonian, provide analytically tractable platforms where the influence of coupling anisotropy, external fields, and thermal effects on coherence and entanglement can be understood in detail \cite{Amico2008, Werlang2010}. Furthermore, two-qubit $XY$ systems are widely used as benchmark models for testing quantum control protocols, studying decoherence, and validating quantum hardware implementations \cite{Barreiro2011, Blais2021}. {The insights gained at the two-qubit level provide an essential foundation for extending the analysis to many-body systems while maintaining physical transparency and interpretability.}

	We focus on four key quantifiers: concurrence $C$, local quantum uncertainty (LQU), Bell--CHSH nonlocality ($\mathbb{B}$), and the $l_1$-norm of quantum coherence ($C_l$). {These quantities capture different aspects of quantum correlations in a two-qubit system.} Concurrence measures bipartite entanglement. LQU is a discord-type measure, and can detect quantum correlations even in separable states \cite{Adesso}. Bell nonlocality tests whether the system can be described by local hidden-variable models \cite{Horodecki1995}. Coherence measures the superposition content that underlies quantum resources \cite{Coherence}.
	{Together, these four quantities provide a compact and informative set of indicators.} They are theoretically distinct and experimentally relevant. They are also sufficient to describe the correlation structure of a two-qubit state without unnecessary redundancy. Other quantities, such as quantum discord \cite{Ollivier2001}, measurement-induced nonlocality \cite{Luo2011}, or the relative-entropy coherence \cite{Coherence}, could also be considered. {However, they provide information similar to that already captured by the above measures.} They do not qualitatively change the behavior observed in our system. For this reason, we restrict our analysis to this widely used set of quantifiers.

	We study quantum correlations and coherence in a thermal Heisenberg $XY$ system under a magnetic field $B$. The system includes tunable parameters: magnetic anisotropy $\delta_m$, coupling anisotropy $\delta_c$, and Dzyaloshinskii-Moriya interaction $D$. 
	To understand how the system behaves quantum mechanically, we use different quantifiers: concurrence ($C$), local quantum uncertainty (LQU), Bell-CHSH observable ($\mathbb{B}$), and the $l_1$-norm ($C_l$).
	Our results reveal that quantum effects are strongest at low temperatures and  and {within suitable ranges of} the magnetic field. When $\delta_m$ is small, entanglement ($C$) drops suddenly as temperature increases. For larger $\delta_m$, the drop is smoother, meaning that  magnetic anisotropy helps protect quantum states. Strong $\delta_m$ also reduces local correlations (LQU). Nonlocality ($\mathbb{B}$) reaches a peak at some magnetic field value but disappears quickly as temperature increases. Bell-CHSH violations ($\mathbb{B} > 2$) only survive at low temperatures. Coherence ($C_l$) is the most robust, and lasts even at higher temperatures, especially for large $\delta_m$. The DM interaction ($D$) is necessary to generate entanglement. We find that as long as the temperature increases, quantum features disappear one by one. First, nonlocality ($\mathbb{B}$) disappears, then entanglement ($C$), and finally, local correlations (LQU). Coherence ($C_l$) is the most resistant, and remains for the longest time. These results can help design spin-based quantum devices that keep their quantum properties in thermal environments.
	
	The present paper is organized as follows. In Sec.~\ref{sec2}, we consider  the Hamiltonian of an anisotropic two-qubit $XY$ Heisenberg spin-$\frac{1}{2}$ chain model with Dzyaloshinskii--Moriya interaction, {magnetic anisotropy,} and an external magnetic field. We determine the {energy spectrum} and compute the corresponding thermal density {matrix}. In Sec.~\ref{sec3}, we analyze quantum correlations and coherence by examining the effect of magnetic anisotropy on four quantum quantifiers. Specifically, after presenting the theoretical formulations, we discuss $C$, LQU, $\mathbb{B}$, and $C_l$ numerically. 
	{Sec.~\ref{V} presents a comparison between our results and previous studies.}
	In Sec.~\ref{sec5}, we numerically investigate the hierarchy among quantum correlations and coherence measures in our system. Finally, we summarize, and conclude {the main results}.
	

	\section{Energy spectrum and thermal density \label{sec2}}
	
	\subsection{Energy spectrum}

	We consider {a two-qubit anisotropic Heisenberg $XY$ model} in which the spins-$\frac{1}{2}$ interact through an exchange coupling, and are subjected to an external magnetic field $B$ applied along the $z$-axis. {This model captures the essential physics of spin-based quantum systems, and provides a simple but effective framework for investigating quantum correlations and coherence in the presence of external perturbations.} The behavior of the system is governed by the following Hamiltonian:
	%
	%
	\begin{equation}
		H = B_{1z}\sigma_{1}^{z} + B_{2z}\sigma_{2}^{z}
		+ J_x\,\sigma^{x}_1\sigma^{x}_2 + J_y\,\sigma^{y}_1\sigma^{y}_2
		+ D\left(\sigma_{1}^{x}\sigma_{2}^{y} - \sigma_{1}^{y}\sigma_{2}^{x}\right).
	\end{equation}
	Here, the parameters \(B_{\alpha z}\), \(J_\alpha\) (\(\alpha = 1,2\)), and \(D\) are taken to be dimensionless. {This choice follows} from normalizing the physical Hamiltonian according to
	$
	H = \frac{H^{\text{phys}}}{E_0},
	$
	where \(H^{\rm phys}\) carries the physical units and \(E_0\) denotes a characteristic energy scale of the system. {The Hamiltonian can then be written as}
	%
	\begin{widetext}
	\begin{align}
		H=
		B	(1+\delta_{m})\sigma_{1}^{z}+
		B(1-\delta_{m})\sigma_{2}^{z}+J(\sigma_{1}^{+}\sigma_{2}^{-}+\sigma_{1}^{-}\sigma_{2}^{+})+ J \delta_{c} (\sigma_{1}^{+}\sigma_{2}^{+}+\sigma_{1}^{-}\sigma_{2}^{-}) 
		+ 
		D	\left(\sigma_{1}^{x}\sigma_{2}^{y}-\sigma_{1}^{y}\sigma_{2}^{x}\right) \label{(1)},
	\end{align}
	\end{widetext}
	where $\sigma_{\alpha}^{\pm}$ ($\alpha = 1,2$) are the spin raising and lowering operators. The coupling parameters are defined as
	$J=\frac{1}{2}(J_{x}+J_{y})$ and $\delta_{c}=\frac{J_{x}-J_{y}}{J_{x}+J_{y}}$, with $\delta_{c}\in[-1,1]$.  
	The magnetic field is given by $B=\frac{1}{2}(B_{1z}+B_{2z})$, while the magnetic anisotropy is defined as $\delta_{m}= \frac{B_{1z}-B_{2z}}{B_{1z}+B_{2z}}$, with $\delta_{m}\in[0,1]$, such that {$\delta_m=0$ corresponds to the isotropic case}. The last term represents the Dzyaloshinskii--{Moriya} (DM) interaction between the two spins \cite{Dz,Morya}. By introducing the spin basis $\mathcal{B}_{c}=\left\lbrace \mid00\rangle,\mid01\rangle,\mid10\rangle,\mid11\rangle \right\rbrace$, we can write the Hamiltonian as
	%
	%
	%
	%
	\begin{align}
		H =
		\begin{pmatrix}
			-{2}B & 0 & 0 & J\delta_c \\
			0 & {2}B\delta_m & J - iD & 0 \\
			0 & J + iD & -{2}B\delta_m & 0 \\
			J\delta_c & 0 & 0 & {2}B
		\end{pmatrix}
		\label{2}.
	\end{align}
	From the relation $\det\left(H- E\mathbb{I}_4\right)=0$,  with  $\mathbb{I}_4$ is the matrix unit, we show that   the four eigenenergies of $H$ take the expressions
	\begin{align}
		E_{1}&=+\sqrt{J^{2}+D^{2}+{4}\delta_{m}^{2}B^{2}}\label{E1},\\
		E_{2}&=-\sqrt{J^{2}+D^{2}+{4}\delta_{m}^{2}B^{2}},\label{E2}\\
		E_{3}&= +\sqrt{J^{2}\delta_{c}^{2}+{4}B^{2}},\label{E3}\\
		E_{4}&= -\sqrt{J^{2}\delta_{c}^{2}+{4}B^{2}}.\label{E4}
	\end{align}
	The associated normalized eigenvectors can be obtained using the eigenvalue equation $(H\ket{\phi}= E\ket{\phi})$ to arrive at the four components, respectively,
	\begin{align}
		&	|\phi_{1}\rangle=N_{1}^{+}\left(\frac{\sqrt{J^{2}+D^{2}}}{E_{1}- {{2}B\delta_{m}}}  e^{i\xi}|01\rangle+|10\rangle\right)\label{st1}, \\
		&	|\phi_{2}\rangle=N_{1}^{-}\left(-\frac{\sqrt{J^{2}+D^{2}}}{E_{2}+ {{2}B\delta_{m}}}  e^{i\xi}|01\rangle+|10\rangle\right), \label{st2}\\
		&	|\phi_{3}\rangle=  N_{2}^{+}\left( |00\rangle + \frac{J\delta_{c}}{E_{3}- {{2}B}}|11\rangle\right), \label{st3}\\
		&	|\phi_{4}\rangle=  N_{2}^{-}\left( |00\rangle -\frac{J\delta_{c}}{E_{4}+ {{2}B}}|11\rangle\right), \label{st4}
	\end{align}
	where the normalization constants and angle are
	\begin{align}
		&	N_{1}^{\pm}=\frac{E_{1,2}\mp {{2}\delta_{m}B} }{\sqrt{J^{2}+D^{2}+(E_{1,2}\mp{{2}B\delta_{m}})^{2}}},\\
		&	N_{2}^{\pm}=\frac{E_{3,4}\mp {{ 2}B}}{\sqrt{J^{2}\delta_{c}^{2}+(E_{3,4}\mp {{2}B})^{2}}},\\
		&	\xi= \arctan \frac{D}{J}.
	\end{align}

	\subsection{Thermal density}
	
	Within the framework of a system in thermal equilibrium at temperature $T$, its statistical properties are fully described by the thermal (canonical) density operator. This operator reflects the mixed-state nature of the system arising from thermal fluctuations and is expressed as \cite{Li2008}
	\begin{align}
		\rho(T) = \frac{1}{\mathcal{Z}} e^{-\beta H},
	\end{align}
	where $\mathcal{Z} = \mathrm{Tr}(e^{-\beta H})$ is the partition function. 
	Here, $\beta = 1/T$, and  $T$ being the dimensionless temperature defined through the normalization
	$
	T = \frac{k_B T^{\rm phys}}{E_0},
	$
	with $T^{\text{phys}}$ being the physical temperature, and $E_0$ is the characteristic energy scale used to make all Hamiltonian parameters dimensionless. 
	For the anisotropic two-qubit Heisenberg $XY$ model considered here, the thermal density matrix $\rho(T)$ can be written explicitly as
	\begin{align}
		\rho(T) = \frac{1}{\mathcal{Z}} \sum_{j=1}^{4} e^{-\beta E_j} |\phi_j\rangle \langle \phi_j|,
	\end{align}
	where the eigenvalue $E_j$ (\ref{E1}-\ref{E4}),  and the eigenstates $|\phi_j\rangle$ (\ref{st1}-\ref{st4}) are corresponding to the   Hamiltonian \eqref{2}.
	%
	%
	%
	%
	%
	%
	%
	After straightforward algebra, we show that  the  partition function is
	\begin{widetext}
	\begin{align}\label{pafu}
		\mathcal{Z}=2\left[ \cosh\left(\frac{\sqrt{J^{2}+D^{2}+{ 4}\delta_{m}^{2}B^{2}}}{T} \right)+ \cosh\left(\frac{\sqrt{J^{2}\delta_{c}^{2}+{4}B^{2}}}{T} \right)\right].
	\end{align}
	\end{widetext}
	Consequently, using (\ref{E1}-\ref{st4}) together with \eqref{pafu}, we obtain the thermal density 
	\begin{align}
		\rho(T)= \frac{1}{\mathcal{Z}} \begin{pmatrix}
			\rho_{11}& 0  & 0 & \rho_{14} \\ 
			0&\rho_{22}& \rho_{23} & 0 \\ 
			0& \rho_{23}^{\ast} & \rho_{33} &0  \\ 
			\rho_{14}&0 & 0 &\rho_{44} 
		\end{pmatrix}  \label{density},
	\end{align}
	where the matrix elements are given by
	\begin{widetext}
	\begin{align}
		&	\rho_{11}= (N_{2}^{+})^2e^{-\beta E_{3}}+(N_{2}^{-})^2e^{\beta E_{3}},\label{rho11}\\
		&	\rho_{14}= J\delta_{c}\left[- \frac{(N_{2}^{+})^2 }{E_{3}-{ 2}B}e^{-\beta E_{3}} + \frac{(N_{2}^{-})^2 }{E_{3}+{2}B}e^{\beta E_{3}}\right],\\
		&	\rho_{22}=\left(J^{2}+D^2\right)\left[ \frac{(N_{1}^{+})^2 }{(E_{1}-{2}\delta_{m}B)^{2}}e^{-\beta E_{1}}+\frac{(N_{1}^{-})^2 }{(E_{1}+{2}\delta_{m}B)^2}e^{\beta E_{1}} \right],\\
		&	\rho_{23}=\sqrt{J^{2}+D^2}e^{i\xi}\left[ \frac{(N_{1}^{+})^2 }{E_{1}-{2}\delta_{m}B}e^{-\beta E_{1}}-\frac{(N_{1}^{-})^2 }{E_{1}+{ 2}\delta_{m}B}e^{\beta E_{1}}\right],\\
		&	\rho_{33}=(N_{1}^{+})^2e^{-\beta E_{1}}+(N_{1}^{-})^2e^{\beta E_{1}},\\
		&	\rho_{44}= (J\delta_{c})^2\left[\frac{(N_{2}^{+})^2}{(E_{3}-{ 2}B)^{2}}e^{-\beta E_{3}}+\frac{(N_{2}^{-})^2}{(E_{3}+{2}B)^{2}} e^{\beta E_{3}}\right]. \label{rho44}
	\end{align}
	\end{widetext}
	Here,  the asterisk $({}^*)$ indicates complex conjugation.
	It should be noted that the normalization condition can  easily be verified 
	\begin{align}
		\sum_{j=1}^{4} \rho_{jj} = \mathcal{Z},
	\end{align}
	which ensures that the thermal density matrix is properly normalized. This relation confirms the internal consistency of our formulation, since the trace of the density matrix must be equal to unity in the canonical ensemble. Moreover, the explicit expressions obtained for the matrix elements of $\rho(T)$ define a compact structure, commonly known as an X-state, which is particularly convenient for the analytical evaluation of various quantum correlation measures.
	

	\section{Quantum correlations and Coherence \label{sec3}}

	Here, the term critical is used to denote parameter values at which quantum correlations, coherence, or other observables undergo abrupt changes.   Although, these points do not correspond to many-body phase transitions, they serve as convenient markers for controlling quantum behavior in the two-qubit system.
	
	\subsection{Entropy of formation}
	
	Many entanglement measures have been proposed and extensively studied in the literature. These measures are used to characterize the quantum correlations between different parts of a composite system. Among them, the entropy of formation $E_F$ \cite{Woo} is a fundamental and widely used quantity, especially for bipartite two-qubit systems. It quantifies the minimal amount of entanglement required to prepare a given quantum state by means of pure entangled states and local operations assisted by classical communication. For a general mixed state of two qubits, the entropy $E_F$ is defined as \cite{Woo}
	\begin{eqnarray}
		E_{F}=h\left(\frac{1}{2}+\frac{1}{2}\sqrt{1-C^{2}}\right),
	\end{eqnarray}
	where $h(x)=-x\log_{2}(x)-(1-x)\log_{2}(1-x)$ is the binary entropy function, and $C$ denotes the concurrence,
	\begin{eqnarray}
		C[\rho](T)=\max\left[0,2\lambda_m(R)-\tr(R)\right],
	\end{eqnarray}
	where $\lambda_{m}(R)$ denotes the largest eigenvalue of the matrix $R$, which is defined as
	\begin{eqnarray}
		R=\rho\left(\sigma_{1}^{y}\otimes\sigma_{2}^{y}\right)\rho^{\star}\left(\sigma_{1}^{y}\otimes\sigma_{2}^{y}\right).
	\end{eqnarray}
	It is worth noting that $C=0$ for a separable state, whereas $C=1$ for a maximally entangled state.
	Using the thermal density matrix~(\ref{density}), and after straightforward calculations, the concurrence can be written as
	\begin{align}
		C[\rho](T)=\tfrac{2}{\mathcal{Z}}\max\left[ 0,|\rho_{23}|-\sqrt{\rho_{11}\rho_{44}},|\rho_{14}|-\sqrt{\rho_{22}\rho_{33}}\right].\label{concur}
	\end{align}
	We first make the following remarks. Indeed, it is important to note that the two symmetries $C(-J)=C(J)$ and $C(\delta_{c})=C(-\delta_{c})$ are satisfied. Therefore, for numerical purposes, we consider only the cases $\delta_{c}\in[0,1]$ and $J>0$. Moreover, for the isotropic case ($\delta_{c}=0$),~\eqref{concur} reduces to
	\begin{eqnarray}
		C[\rho](T)=\tfrac{2}{\mathcal{Z}}|\rho_{23}|.
	\end{eqnarray} 
	
	\begin{figure*}
		\centering
		\subfloat[]{
			\includegraphics[scale=0.38]{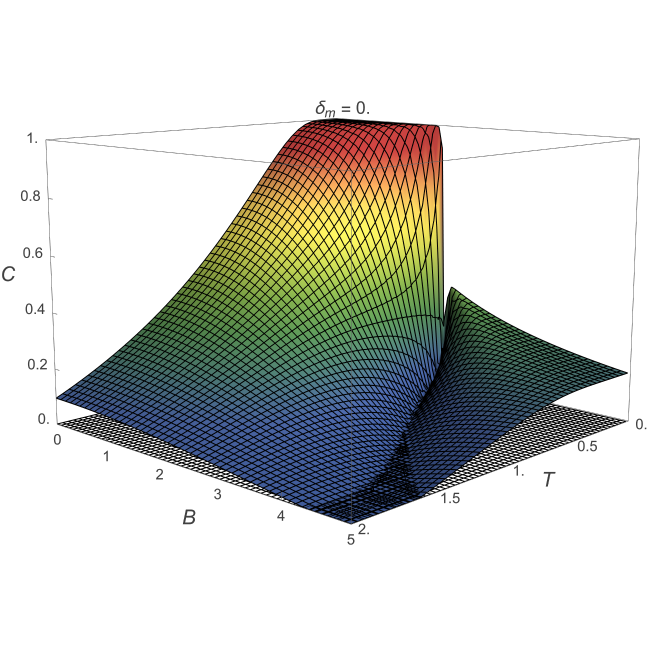}}
		\subfloat[]{\includegraphics[scale=0.38]{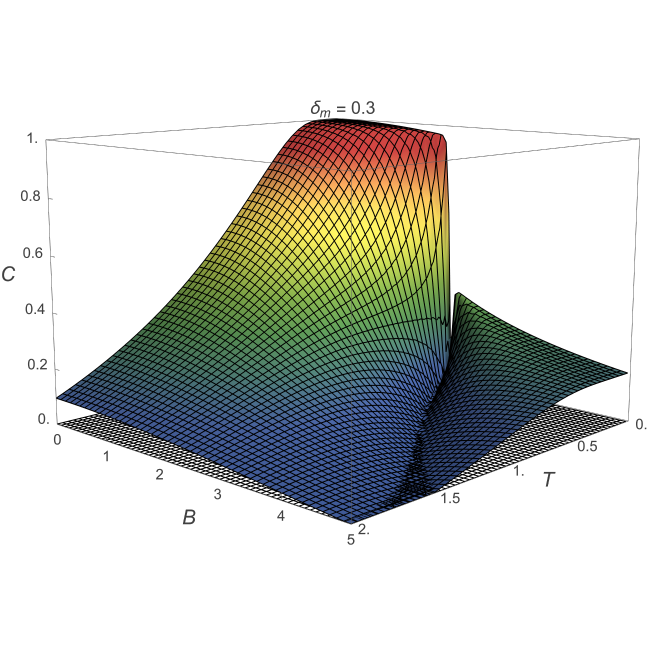} }
		\subfloat[]{\includegraphics[scale=0.38]{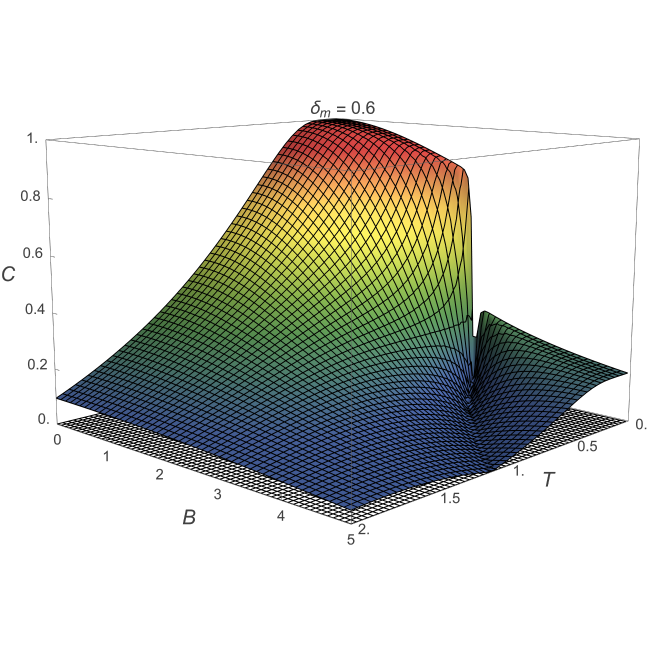}} 
		\subfloat[]{	\includegraphics[scale=0.38]{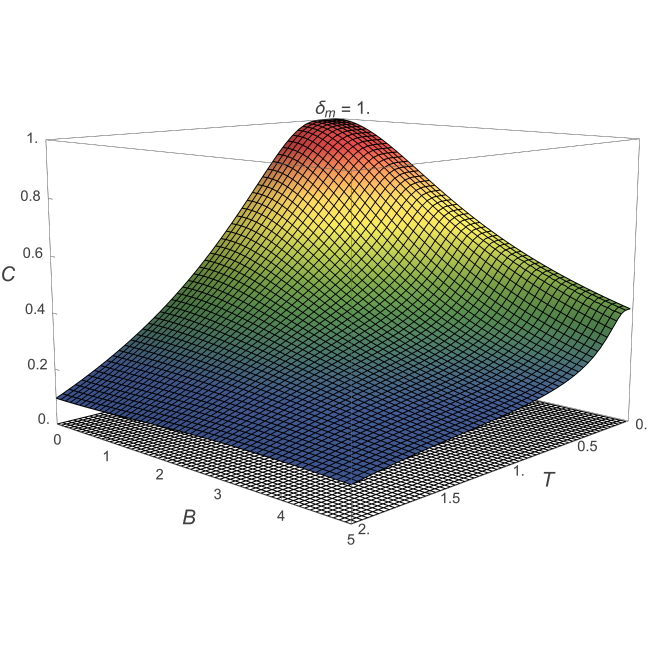}}
		\caption{Concurrence $C$ as a function of temperature $T$, and magnetic field $B$ for $\delta_{c}=0.9$, $J=1$, and $D=2$. Panels (a)–(d) correspond to $\delta_{m}=0$, $0.3$, $0.6$, and $1$, respectively.}
	\label{fig1} 
\end{figure*}

In Figure~\ref{fig1}, we plot the quantum entanglement, quantified by the concurrence $C$, as a function of temperature $T$m and magnetic field $B$ for the parameters $(\delta_c = 0.9,\; J = 1,\; D = 2)$, and  different values of the magnetic anisotropy $\delta_m$.
The plots show that concurrence reaches its largest values at low temperatures and for suitable magnetic fields. It then decreases monotonically as the temperature increases. This behavior is consistent with thermal decoherence in quantum spin systems.
For $\delta_m = 0$, the entanglement exhibits sudden death around $B \approx 4.1$ and $T \approx 1.6$. At this point, the concurrence drops abruptly to zero. As $\delta_m$ increases, this sudden-death behavior becomes less pronounced. The entanglement then survives over a wider range of $T$ and $B$. For $\delta_m = 1$, the entanglement is more robust and decays smoothly rather than collapsing abruptly.
These results indicate that increasing $\delta_m$ enhances the ability of the system to maintain quantum correlations under external perturbations. Therefore, magnetic anisotropy can play a key role in stabilizing the entanglement required for quantum information applications.

\subsection{Local quantum uncertainty}

To clarify the difference between quantum and classical correlations, consider the following example. Classical correlations arise from shared statistical information. For example, two coins are more likely to show the same face due to their preparation conditions. In contrast, quantum correlations reflect intrinsic non-classical features of a quantum state that cannot be explained by probabilities alone. Measures such as local quantum uncertainty (LQU) capture these nonclassical correlations, going beyond what classical correlations or entanglement quantify. LQU is a legitimate discord-type quantifier of nonclassical correlations that vanishes if and only if the state has zero quantum discord (QD) \cite{Adesso, Ollivier2001}.
A standard example is provided by a two-qubit separable state that has non-zero quantum discord. Such a state exhibits correlations that are not captured by entanglement but are revealed by LQU, demonstrating purely quantum correlations in the absence of classical explanations \cite{Ollivier2001}.
LQU is particularly appealing for two reasons. First, it is a useful tool for analyzing quantum correlations in many different systems. Second, it is computationally efficient, because it can be obtained directly from the eigenvalues of a 
$	3\times 3$
symmetric matrix, unlike quantum discord which requires an optimization procedure.
%
%
The matrix $\mathcal{M}(\rho)$ is given by 
\cite{Adesso}
\begin{align}
	\mathcal{M}_{ij}(\rho)=\tr\left[\sqrt{\rho}\left(\sigma_{i}^{A}\otimes \mathcal{I}^{B} \right)\sqrt{\rho} \left(\sigma_{j}^{A}\otimes \mathcal{I}^{B}\right) \right]\label{M},
\end{align}
where $\mathcal{I}^{A,B}$ denote the $2 \times 2$ identity matrices, and $\sigma_j$ ($j=1,2,3$) are the Pauli matrices.


%

To proceed further, we introduce a unitary operator \(U\) in order to transform the thermal density matrix in~(\ref{density}) into a more suitable representation. This transformation simplifies the analysis and makes the derivation of the relevant quantities more transparent. We thus obtain
%
%
\begin{align}
	\tilde\rho(T) = U \rho(T) U^\dagger,
\end{align}
where \(U\) is a diagonal matrix in the computational basis \(\{|00\rangle, |01\rangle, |10\rangle, |11\rangle\}\)
\begin{widetext}
\begin{align}
	U = \text{diag}
	\left(e^{-i(\phi_{14} + \phi_{23})},\ e^{-i(\phi_{14} - \phi_{23})},\ e^{i(\phi_{14} - \phi_{23})},\ e^{i(\phi_{14} + \phi_{23})}\right),
\end{align}
\end{widetext}
such that $\phi_{14}$, and $\phi_{23}$ are the arguments of the matrix elements $\rho_{14}$, and $\rho_{23}$, respectively. After this transformation, the density matrix becomes
%
%
%
\begin{align}\label{transformed density}
	\tilde\rho(T) = \frac{1}{\mathcal{Z}}
	\begin{pmatrix}
		\rho_{11} & 0 & 0 & |\rho_{14}| \\
		0 & \rho_{22} & |\rho_{23}| & 0 \\
		0 & |\rho_{23}| & \rho_{33} & 0 \\
		|\rho_{14}| & 0 & 0 & \rho_{44}
	\end{pmatrix},
\end{align}
where all off-diagonal elements are now real, and positive (their magnitudes).
%
%
%
%
%
Note that  local quantum uncertainty (LQU) remains invariant under such transformation, namely we have LQU$(\rho)$=LQU$(\tilde{\rho})$. As a result, we can write
\begin{eqnarray}\label{35}
	\text{LQU}(\rho)=1-\max[\mathcal{M}_{11},\mathcal{M}_{22},\mathcal{M}_{33}].
\end{eqnarray}
The relevant matrix elements $\mathcal{M}_{jj}$ ($j=1,2,3$) can be determined as follows. To this end, we consider the matrix \(\sqrt{\tilde{\rho}(T)}\) given by
\begin{align}\label{sqrho}
	\sqrt{\tilde{\rho}(T)} = \frac{1}{\sqrt{\mathcal{Z}}}
	\begin{pmatrix}
		m_{11} & 0 & 0 & m_{14} \\
		0 & m_{22} & m_{23} & 0 \\
		0 & m_{23} & m_{33} & 0 \\
		m_{14} & 0 & 0 & m_{44}
	\end{pmatrix}	,
\end{align}
where the matrix elements can be obtained from~(\ref{density}) as
\begin{align}
	m_{11} &= \dfrac{\rho_{11}+\sqrt{\rho_{11}\rho_{44}-|\rho_{14}|^{2}}}{\sqrt{\rho_{11} + \rho_{44}+2\sqrt{\rho_{11}\rho_{44}-|\rho_{14}|^{2}}}},  \\
	m_{22} &= \dfrac{\rho_{22}+\sqrt{\rho_{22}\rho_{33}-|\rho_{23}|^{2}}}{\sqrt{\rho_{22} + \rho_{33}+2\sqrt{\rho_{22}\rho_{33}-|\rho_{23}|^{2}}}}, \\
	m_{33} &= \dfrac{\rho_{33}+\sqrt{\rho_{22}\rho_{33}-|\rho_{23}|^{2}}}{\sqrt{\rho_{22} + \rho_{33}+2\sqrt{\rho_{22}\rho_{33}-|\rho_{23}|^{2}}}}, \\
	m_{44} &= \dfrac{\rho_{44}+\sqrt{\rho_{11}\rho_{44}-\rho_{14}^{2}}}{\sqrt{\rho_{11} + \rho_{44}+2\sqrt{\rho_{11}\rho_{44}-|\rho_{14}|^{2}}}}, \\
	m_{14} &= \dfrac{|\rho_{14}|}{\sqrt{\rho_{11} + \rho_{44}+2\sqrt{\rho_{11}\rho_{44}-|\rho_{14}|^{2}}}}, \\
	m_{23} &= \dfrac{|\rho_{23}|}{\sqrt{\rho_{22} + \rho_{33}+2\sqrt{\rho_{22}\rho_{33}-|\rho_{23}|^{2}}}}.
\end{align}
By injecting \eqref{sqrho} into (\ref{M}), we obtain
\begin{align}
	\mathcal{M}_{11}&= \frac{2}{\cal{Z}}(m_{33}m_{11}+m_{44}m_{22}+2m_{23}m_{14}),\\
	\mathcal{M}_{22}&= \frac{2}{\cal{Z}} (m_{33}m_{11}+m_{44}m_{22}-2m_{23}m_{14}),\\
	\mathcal{M}_{33}&= \frac{2}{\cal{Z}}\left[1-2(m_{14}^{2}+m_{23}^{2})\right].
\end{align}
With the conditions  $m_{ij}\geq 0$, and  $\mathcal{M}_{11}\geq \mathcal{M}_{22}$, then    $\text{LQU}(\rho)$ \eqref{35}  reduces to 
\begin{eqnarray}
	\text{LQU}(\rho)=1-\max[\mathcal{M}_{11},\mathcal{M}_{33}].
\end{eqnarray}


\begin{figure*}
	\centering
	\subfloat[]{
		\includegraphics[scale=0.5]{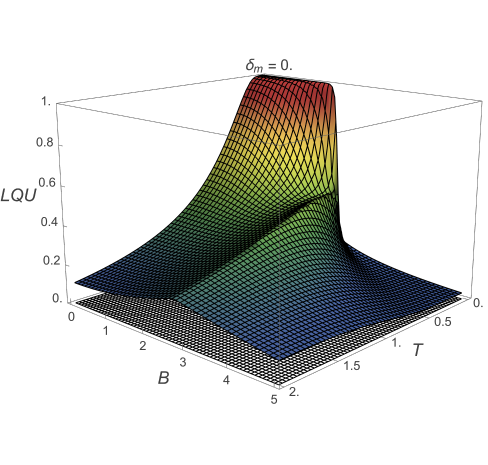}} 
	\subfloat[]{
		\includegraphics[scale=0.5]{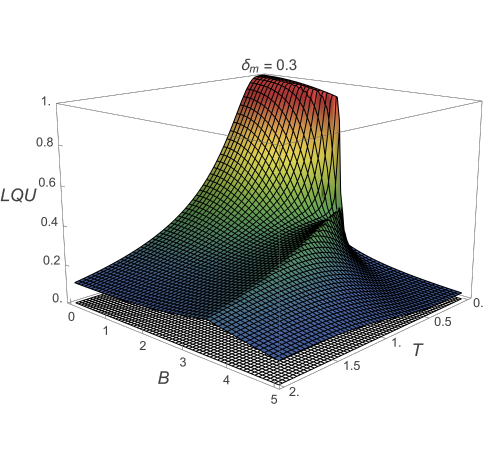}}
	\subfloat[]{
		\includegraphics[scale=0.5]{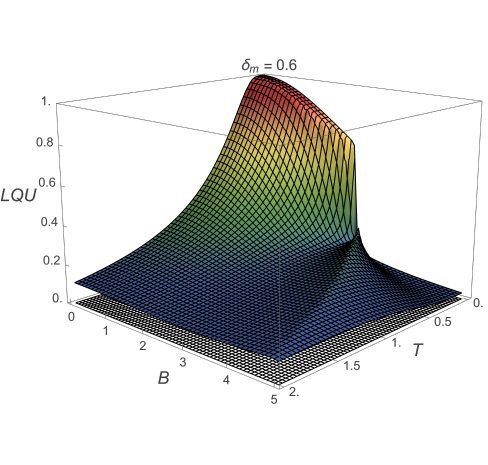}} 
	\subfloat[]{
		\includegraphics[scale=0.5]{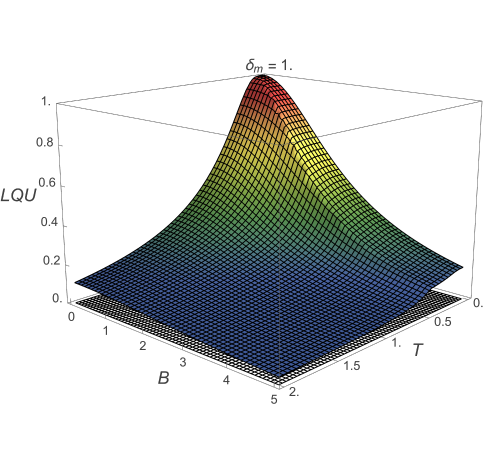}}
	\caption{Local quantum uncertainty (LQU) as a function of temperature $T$, and magnetic field $B$ for $\delta_{c}=0.9$, $J=1$, and $D=2$. Panels (a)–(d) correspond to $\delta_{m}=0$, $0.3$, $0.6$, and $1$, respectively.}
\label{fig3} 
\end{figure*}

Figure \ref{fig3} illustrates the behavior of local quantum uncertainty (LQU) as a function of the magnetic field $B$, temperature $T$, and magnetic anisotropy $\delta_m$. At the origin ($B=0, T=0$), all subfigures exhibit the same LQU value for $0\le\delta_m\le1$, indicating that quantum correlations are insensitive to anisotropy in this specific limit. Consequently, no monotonic dependence of LQU on anisotropy is observed at the null origin. Away from this point, the role of anisotropy becomes evident. For finite magnetic fields and/or temperatures, increasing $\delta_m$ generally leads to a reduction of LQU, reflecting the suppression of quantum correlations induced by stronger anisotropic effects. The temperature dependence of LQU is also regime dependent. For magnetic fields below the critical field $B_c$, LQU decreases with increasing temperature, consistent with thermal decoherence. In contrast, for $B > B_c$, LQU exhibits non-monotonic or weakly varying behavior with temperature, signaling a competition between thermal fluctuations and field-induced quantum correlations. These results demonstrate that the behavior of quantum correlations captured by LQU is governed by the interplay between magnetic anisotropy, temperature, and magnetic field rather than by a simple monotonic trend across the entire parameter space.

\subsection{Bell-CHSH 
nonlocality}

The Bell--Clauser--Horne--Shimony--Holt (CHSH) inequality is a fundamental tool for distinguishing quantum mechanics from local hidden{-}variable theories \cite{Bell,Bell-CHSH,Einstein}. In our model, the X{-}shaped density matrix $\tilde{\rho}(T)$ exhibits quantum correlations that can violate this inequality, thereby demonstrating Bell nonlocality. In this section, we analyze the structure of $\tilde{\rho}(T)$ and derive the conditions under which the CHSH inequality is violated. The density matrix admits the following Bloch decomposition \cite{Schlienz1995}
%
\begin{widetext}
\begin{align}
\tilde{\rho}= \frac{1}{4}\left(\mathcal{I}^{A}\otimes \mathcal{I}^{B}+\vec{r}\cdot\sigma^{A}\otimes \mathcal{I}^{B}+\mathcal{I}^{A}\otimes \vec{s}\cdot \sigma^{B}
\right)	+\frac{1}{4}\sum\limits_{j=1}^{3}c_{j}\sigma^{A}_{j}\otimes\sigma_{j}^{B},
\end{align}
\end{widetext}
where $\vec{r}$, and $\vec{s}$ denote the local Bloch vectors, which are assumed to be aligned along the $z$-axis, i.e., $\vec{r}=r\,\vec{z}$, and $\vec{s}=s\,\vec{z}$, with
\begin{align}
&r=\frac{\rho_{11} +\rho_{22}-\rho_{33}-\rho_{44}}{\mathcal{Z}},\\
&s=\frac{\rho_{11}-\rho_{22}+\rho_{33}-\rho_{44}}{\mathcal{Z}},
\end{align}
while the correlation elements, which encode entanglement, and other nonclassical correlations, are given by
\begin{align}
&	c_{1}=\frac{2}{\mathcal{Z}}(|\rho_{14}|+|\rho_{23}|),\label{C1}\\ 
&	c_{2}=\frac{2}{\mathcal{Z}}(|\rho_{14}|-|\rho_{23}|),\label{C2}\\ &
c_{3}=\frac{\rho_{11}+\rho_{44}-\rho_{22}-\rho_{33}}{\mathcal{Z}}.\label{C3}
\end{align}
By using the Horodecki criterion \cite{Horodecki,Horodecki1}, the maximal violation of the Bell--CHSH inequality for X-states can be written as
\begin{align}
\mathbb{B}[\tilde\rho](T)=2\sqrt{\max\limits_{j<k}[c_{j}^2+c_k^2]}.
\end{align}
From the above considerations,  it takes the form
\begin{align}
\mathbb{B}[\tilde\rho](T)=	 2\max\left[\sqrt{c_{1}^2+c_2^{2}},\sqrt{c_1^{2}+c_3^2}\right].
\end{align}
Using~(\ref{rho11})--(\ref{rho44}), the correlation elements $c_1$, $c_2$, and $c_3$~(\ref{C1})--(\ref{C3}) can be expressed explicitly in terms of the density-matrix elements. It then follows that the $X$-shaped density matrix $\tilde{\rho}$ violates the Bell--CHSH inequality whenever the maximal Bell function $\mathbb{B}$ associated with the correlation matrix satisfies $\mathbb{B} > 2$.



{To place our analysis in the broader context of previous studies on Bell nonlocality in two-qubit systems,} we note that Bell--CHSH violations have been extensively investigated both theoretically and experimentally in {the literature}. {For instance,} a comprehensive review of Bell nonlocality and CHSH inequality tests is given in~\cite{Brunner2014}, where the general framework and the necessary and sufficient conditions for nonlocality in bipartite quantum states are discussed. {On the experimental side,} foundational demonstrations of CHSH violation under strict locality conditions were reported in~\cite{Weihs1998}, and high-fidelity violations with solid-state qubits have also been observed in silicon spin systems \cite{Dehollain2015}. {From a theoretical perspective,} explicit relations between thermal entanglement and maximal Bell inequality violation in two-qubit Heisenberg chains were derived in~\cite{WangZanardi2002}. {In this context,} our work complements these studies by providing analytic expressions for the Bell function $\mathbb{B}$ in terms of the density-matrix elements of $\tilde{\rho}(T)$ for the specific Hamiltonian considered here, and by investigating how the system parameters $J$, $D$, $\delta_m$, $\delta_c$, and $B$ govern the emergence of nonlocal correlations.


\begin{figure*}
\centering
\includegraphics[scale=0.25]{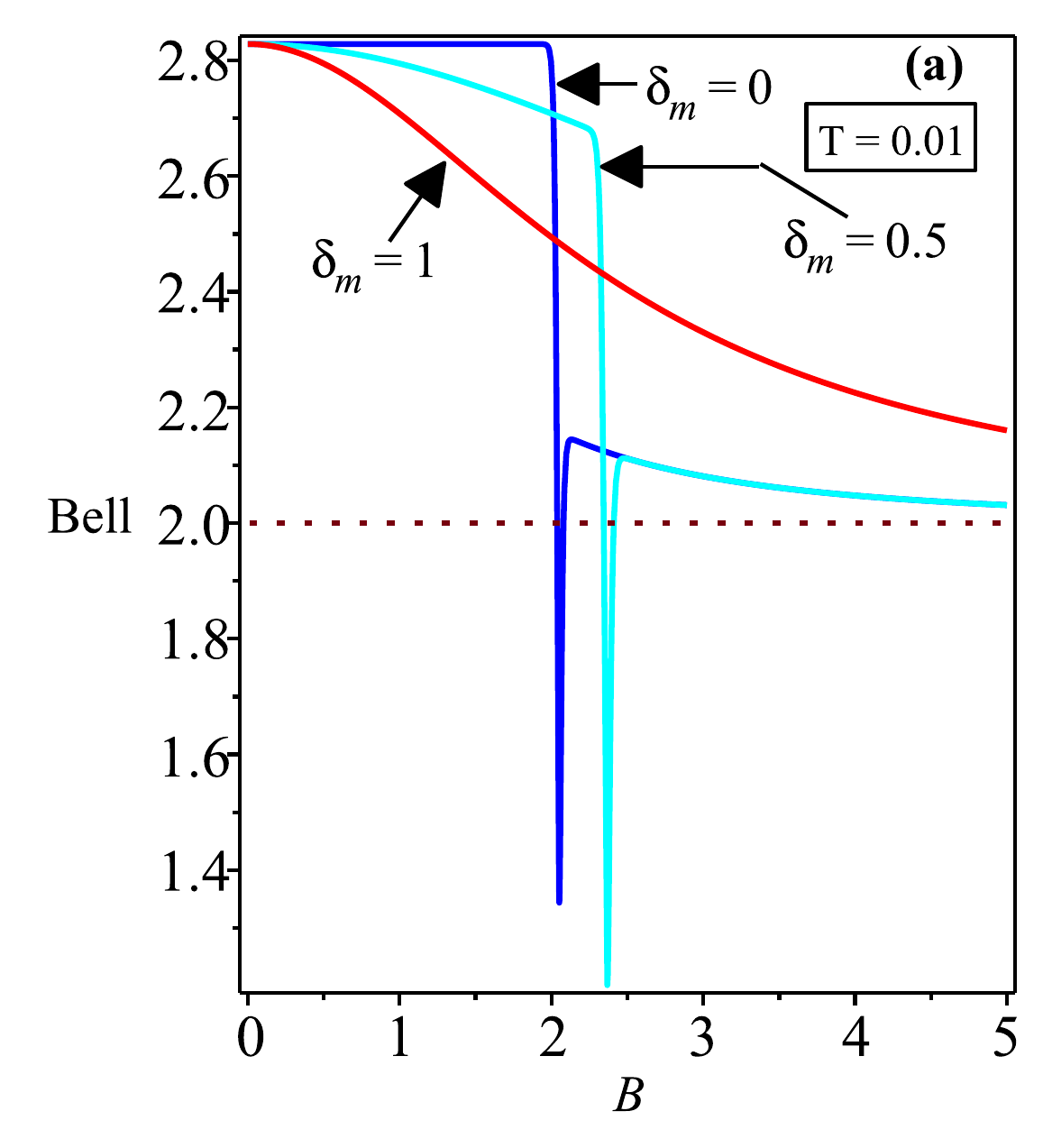} \quad
\includegraphics[scale=0.25]{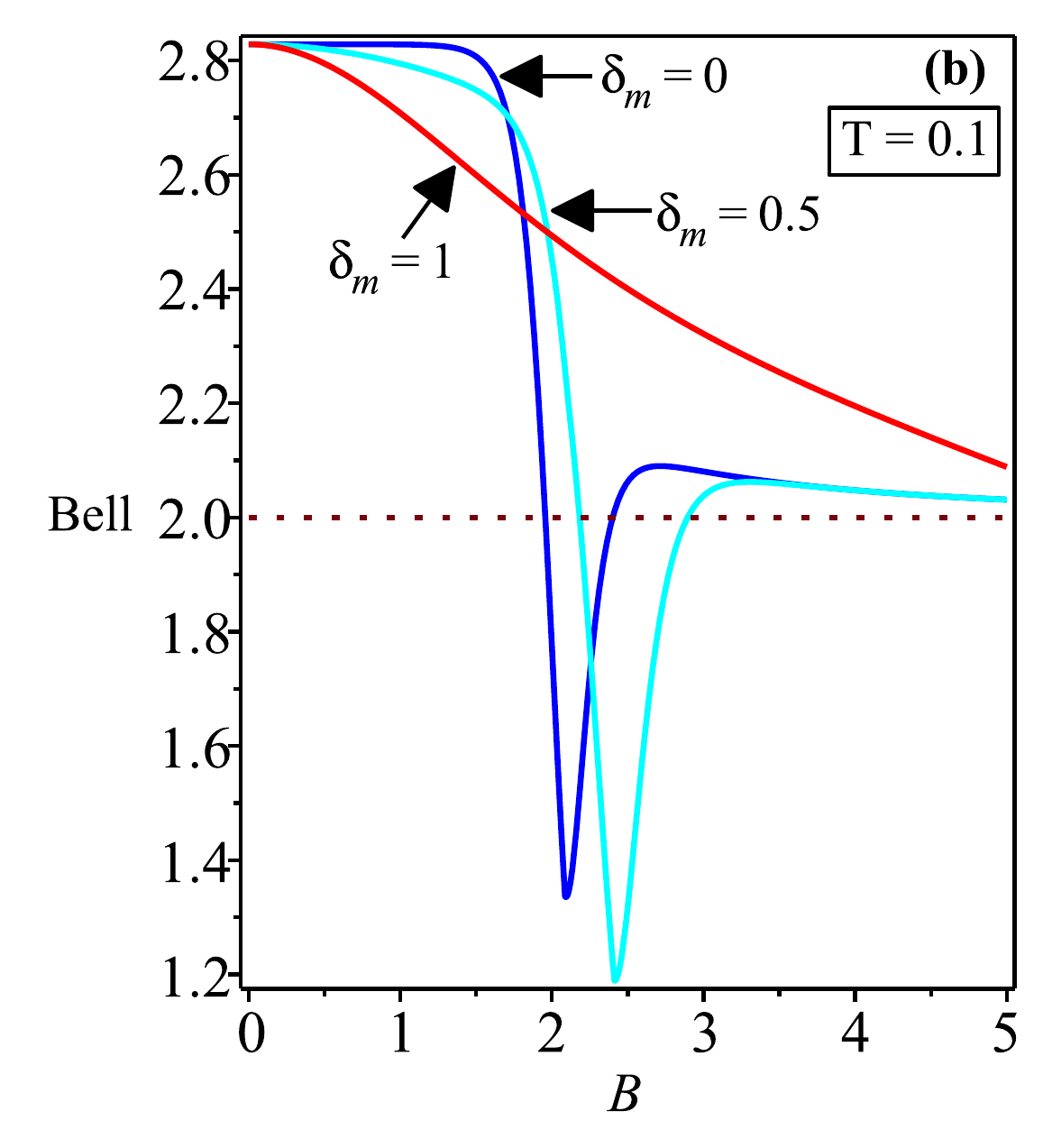} \quad
\includegraphics[scale=0.255]{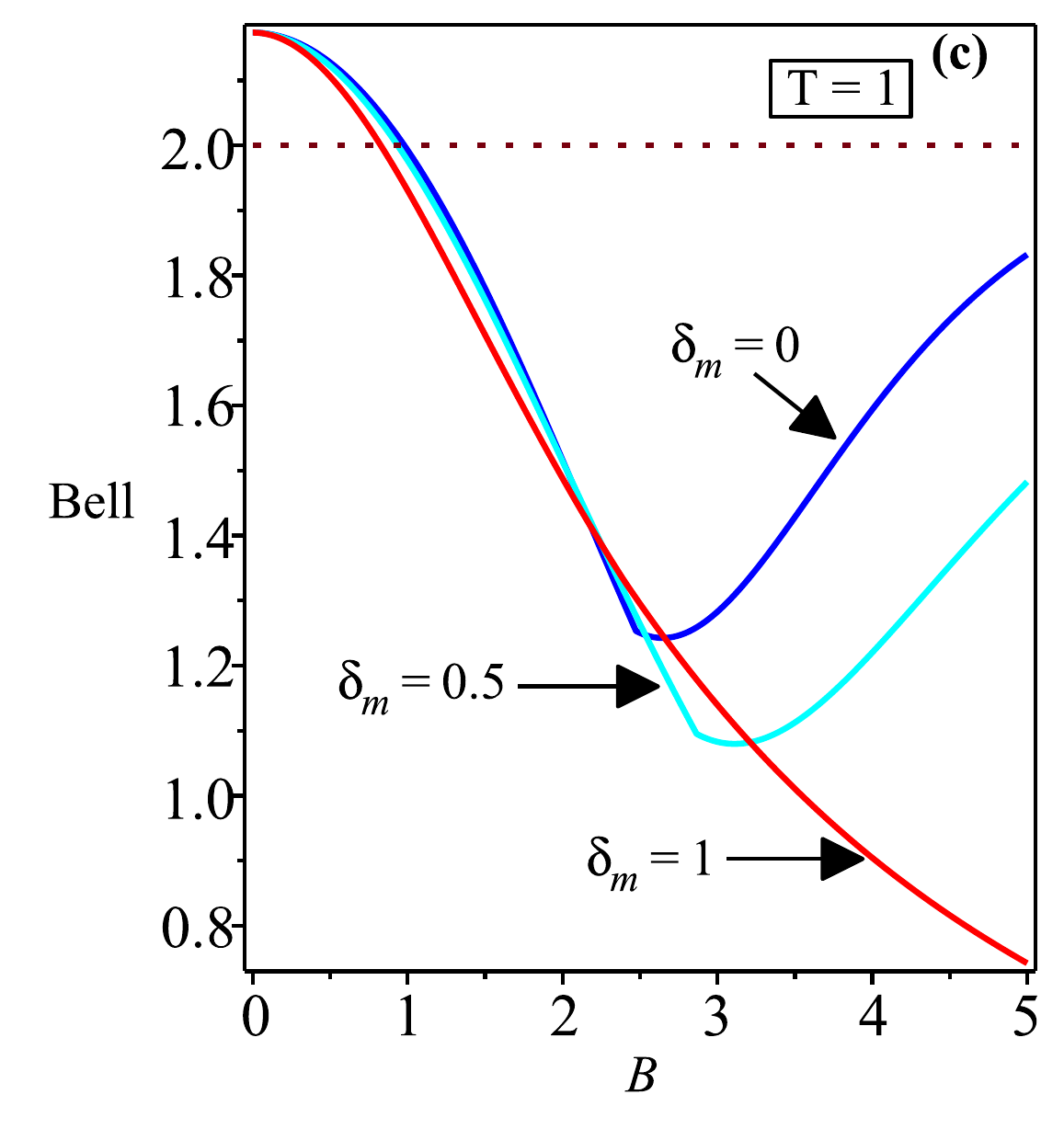} 
\caption{Effect of magnetic anisotropy $\delta_m$, and temperature $T$ on the Bell–CHSH observable $\mathbb{B}$ as a function of the magnetic field $B$ for $D=1$, $J=1$, and $\delta_c = 1$.}
%
\label{fig1B}  
\end{figure*}

In Figure~\ref{fig1B}, we plot the Bell--CHSH observable $\mathbb{B}$ as a function of the magnetic field $B$ for different values of the magnetic anisotropy $\delta_m$, and temperature $T$. The most prominent feature is a sharp drop in $\mathbb{B}$ at a critical field $B_c$. As $\delta_m$ increases, this drop shifts toward higher values of $B$, and $\mathbb{B}$ decreases gradually until it reaches the classical threshold value of $2$. When $\mathbb{B}>2$, the Bell--CHSH inequality is violated, indicating the presence of Bell nonlocality in the system. However, when the temperature is increased to $T=1$, $\mathbb{B}$ falls below $2$ in part of the parameter range.
This behavior shows that thermal effects can weaken and eventually suppress nonlocal quantum correlations. These results demonstrate that both magnetic anisotropy and temperature strongly affect the ability of the system to exhibit Bell nonlocality.

\begin{figure*}
\centering
\includegraphics[scale=0.25]{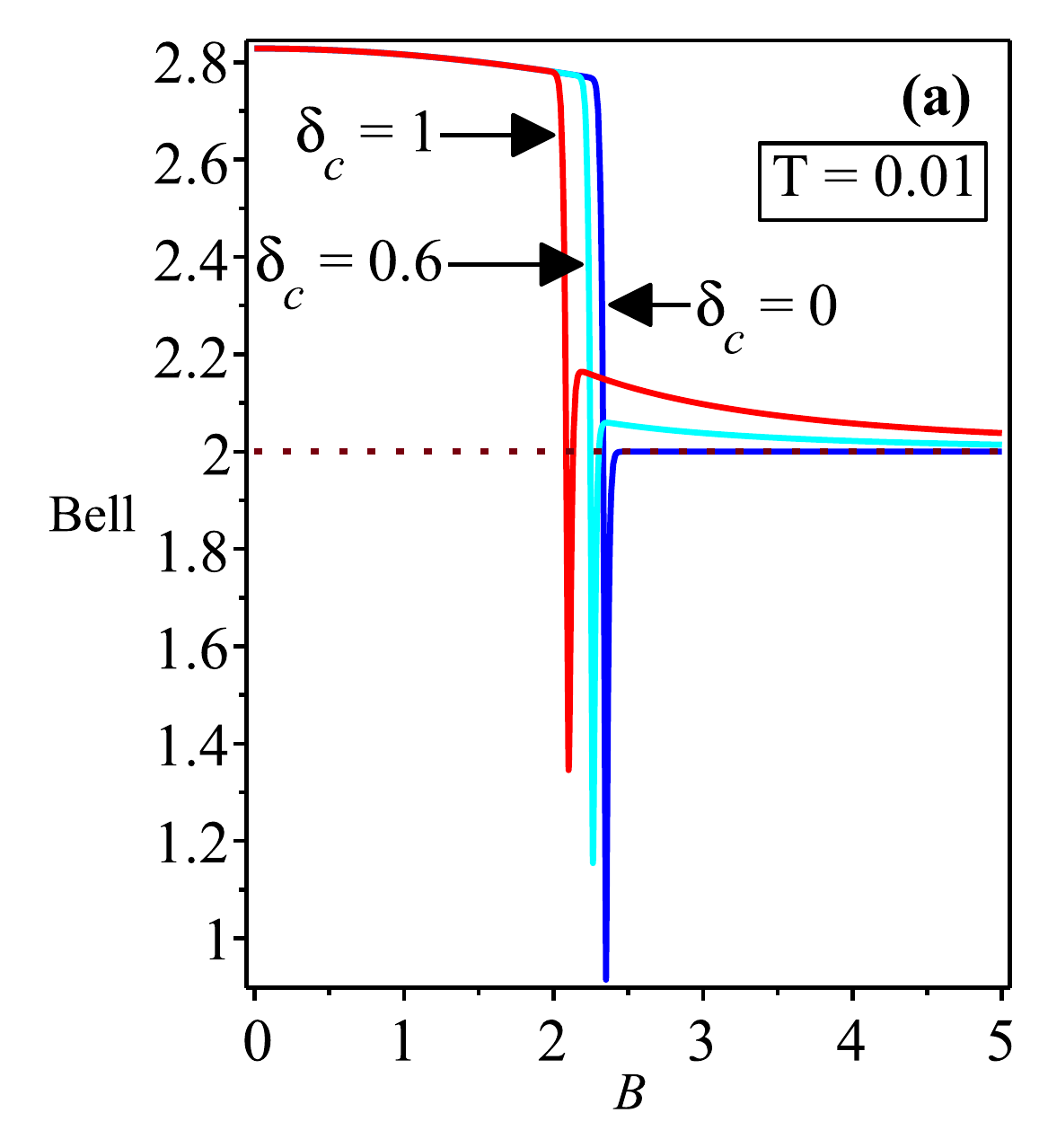} \quad
\includegraphics[scale=0.25]{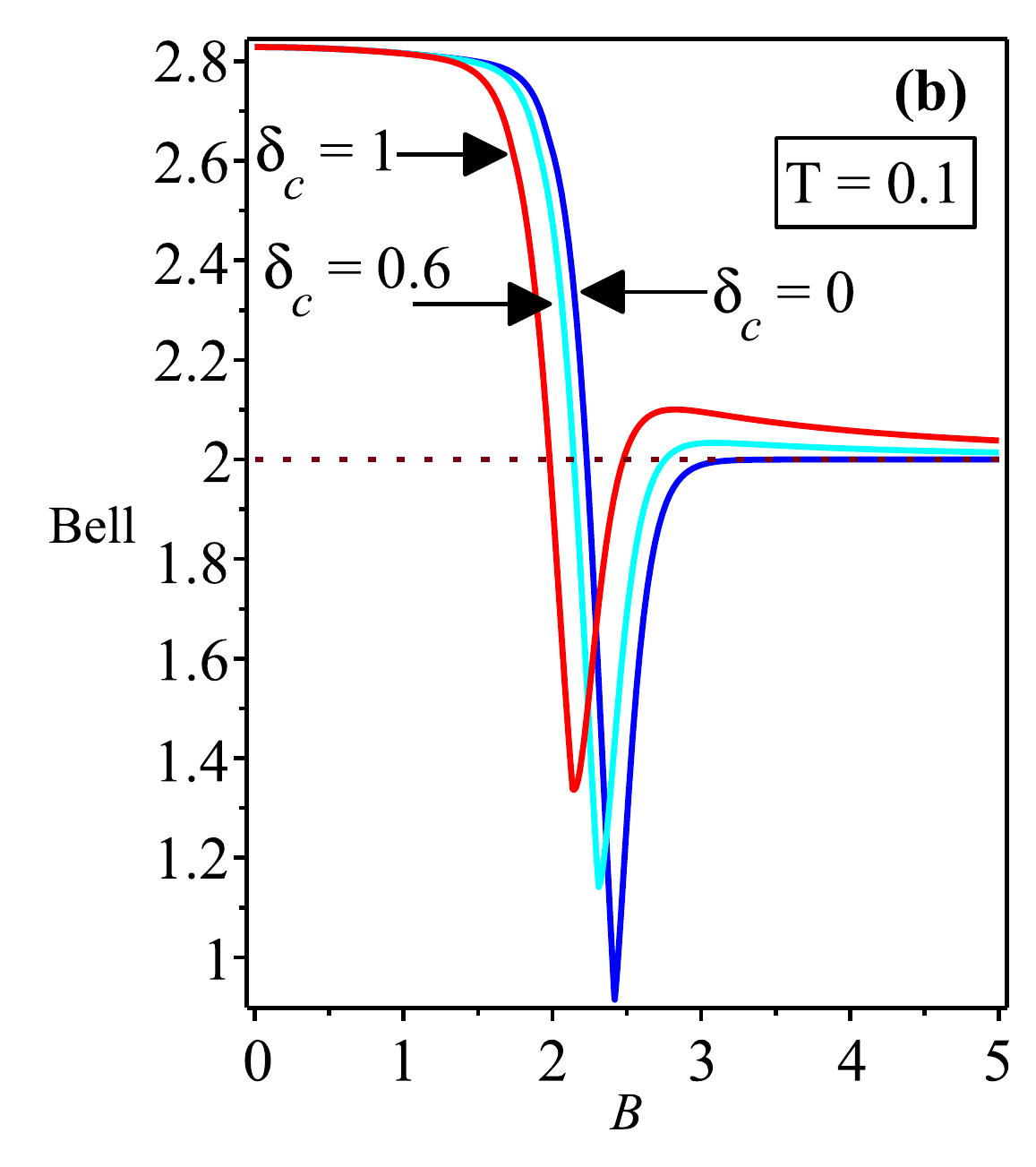} \quad
\includegraphics[scale=0.26]{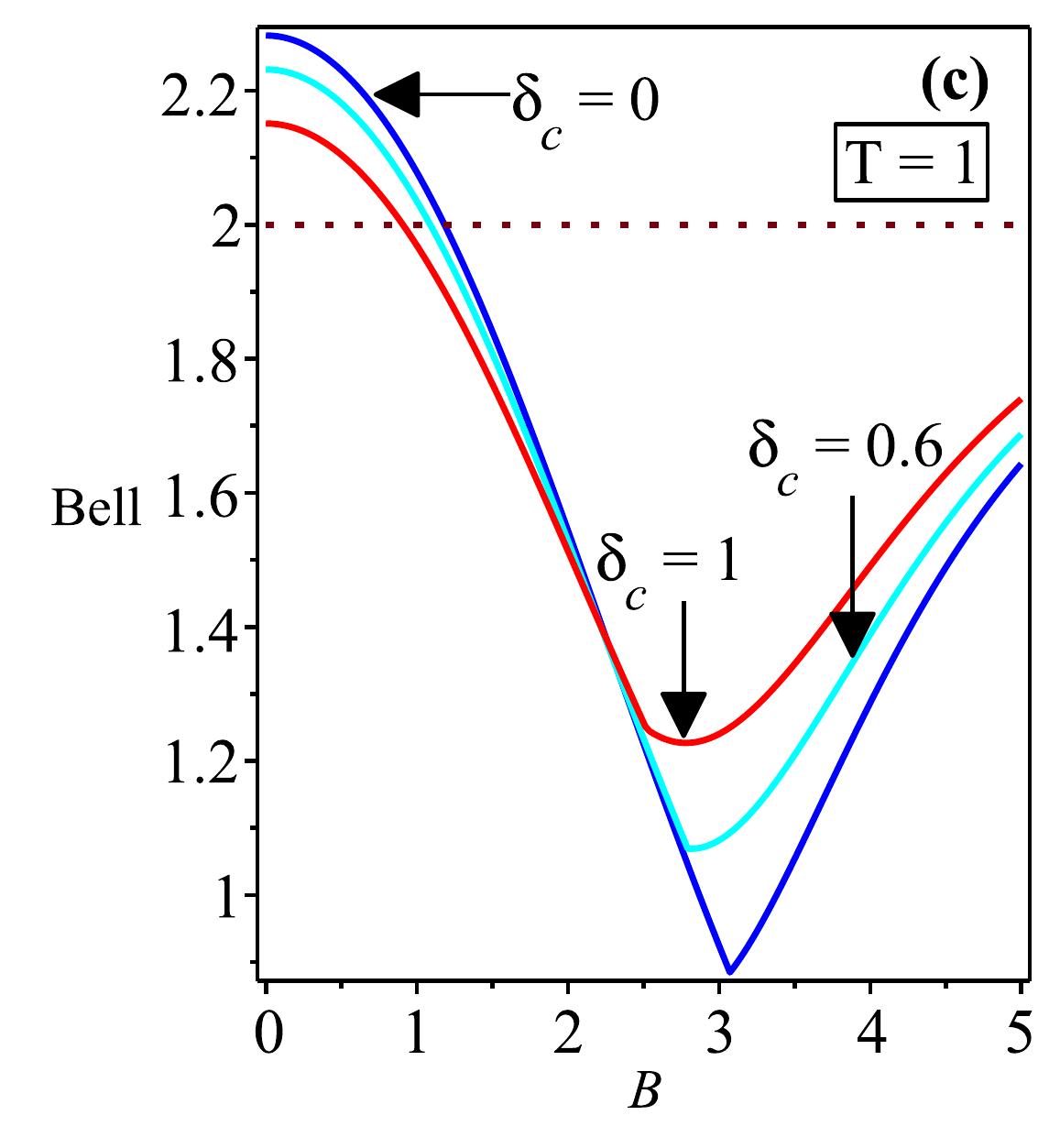} 
\caption{Effect of the coupling anisotropy $\delta_c$, and temperature $T$ on the Bell–CHSH observable $\mathbb{B}$ as a function of the magnetic field $B$ for $D=2$, $J=1$, and $\delta_m = 0.3$.}
\label{fig2B} 
\end{figure*}

Figure \ref{fig2B} {shows} the effect of coupling anisotropy on the Bell-CHSH observable $\mathbb{B}$ as a function of the magnetic field $B$ for $D = 2, J = 1,$ and $\delta_m = 0.3$. Beyond a critical magnetic field, the Bell function increases, indicating that coupling anisotropy enhances quantum nonlocality. In contrast, increasing the temperature $T$ causes the Bell function to decrease and eventually fall below the classical bound of 2. From a physical perspective, the enhancement of the Bell function with coupling anisotropy originates from the strengthening of coherent spin-spin correlations encoded in the off-diagonal elements of the thermal density matrix. The anisotropic coupling favors entangled superposition states, and enhances the correlation coefficients entering the Bell function, particularly at sufficiently strong magnetic fields. On the other hand, thermal fluctuations populate higher-energy states, and drive the system toward a ,ore classical mixed state, thereby suppressing quantum coherence and reducing $\mathbb{B}$. This reveals a clear competition between anisotropy-induced quantum coherence, which promotes Bell nonlocality, and thermal effects, which act as a source of decoherence and {classical behavior}.

The enhancement of Bell nonlocality induced by coupling anisotropy can be traced back to the nature of anisotropic spin-spin interactions. Unlike isotropic couplings, anisotropy breaks rotational symmetry, and selectively amplifies quantum correlations along specific spin axes. This leads to stronger nonclassical correlations encoded in the off-diagonal elements of the density matrix, which directly contribute to the correlation coefficients $c_j$ entering the Bell--CHSH function. {Consequently}, the anisotropic interaction promotes quantum coherence and entangled superposition states rather than classical spin alignment.
{Therefore, the increase in the Bell function} reflects a genuine quantum effect arising from anisotropic interactions rather than an amplification of classical spin-spin correlations.
The suppression of the Bell--CHSH function $\mathbb{B}$ at high temperatures can be understood
as a consequence of thermal mixing rather than environmental decoherence. As the
temperature increases, thermal fluctuations populate multiple energy eigenstates,
including separable or weakly correlated ones, causing the density matrix to approach a
classical mixed state. This thermal mixing reduces the magnitude of the off-diagonal
elements responsible for quantum coherence and directly diminishes the correlation
coefficients
$c_j$ entering the Bell--CHSH inequality. Once thermal effects dominate over
coherent spin--spin interactions, the Bell function $\mathbb{B}$ falls below the classical
bound of $2$, signaling the loss of Bell nonlocality.

\subsection{Quantum coherence}

Quantum coherence is a key component of quantum information processing, and is closely related to quantum interference effects \cite{Coherence}. It serves as a fundamental resource that enables many distinctive quantum features in quantum technologies. In the following analysis, we examine how coherence behaves in our system by using the $l_1$-norm as a quantitative measure. This choice allows us to track the persistence of quantum superposition under various physical conditions, including anisotropy, magnetic field, and temperature. Specifically, quantum coherence is quantified by~\cite{Coherence}
\begin{align}
\mathcal{C}_{l}(\rho) &= \min\limits_{\sigma \in \mathcal{I}} ||\rho - \sigma||,\\
&= ||\rho - \rho_\text{diag}||,\\
&= \frac{1}{\mathcal{Z}} \sum\limits_{i \neq j} |\rho_{ij}|,
\end{align}
where $\mathcal{I}$ denotes the set of incoherent states those for which the coherence vanishes, i.e., $\mathcal{C}_{l}(\sigma) = 0$. Among incoherent states, the closest one to a given quantum state is simply its diagonal in the chosen reference basis.
The $l_1$-norm and the relative entropy of coherence employed in this work were originally introduced within the general framework of quantum coherence \cite{Coherence,Coherence0}. While these measures have been extensively studied in generic spin systems, their behavior under the combined influence of magnetic anisotropy, coupling anisotropy, and Dzyaloshinskii–Moriya interaction remains {less well understood}. The present work extends existing studies by demonstrating that anisotropic interactions can be exploited as effective external control parameters to enhance and stabilize quantum coherence against magnetic-field and temperature-induced decoherence.

Quantum coherence can also be quantified by another widely used measure based on relative entropy. This approach {is defined by minimizing} the relative entropy between the quantum state, and the set of incoherent states. It {measures how far} a quantum state is from its closest incoherent state, providing an information-theoretic view of coherence. This measure is given by
\begin{align}
\mathcal{C}_{R}(\rho) &= \min\limits_{\sigma \in \mathcal{I}} S(\rho || \sigma),\\
&= S(\rho_\text{diag}) - S(\rho),
\end{align}
where $S(\rho) = -\tr(\rho \log \rho)$ is the von Neumann entropy, and $\rho_\text{diag}$ is the diagonal part of $\rho$ in the reference basis. The relative entropy of coherence complements the $l_1$-norm, offering additional insights into coherence in quantum thermodynamics and information theory.
We also point out that both coherence quantifiers are bounded within specific limits, ensuring their consistency across systems of different dimensions \cite{Coherence2}. These bounds take the forms
\begin{align}
&0 \leq \mathcal{C}_{l}(\rho) \leq d - 1,\label{Cl_bound}\\
&0 \leq \mathcal{C}_{R}(\rho) \leq \log_{2} d \label{condition},
\end{align}
where $d$ is the dimension of the Hilbert space associated with the system. For example, for a spin-1/2 system, $d = 2$, so $\mathcal{C}_l$ ranges from 0 to 1 and $\mathcal{C}_R$ from 0 to 1 bit. The lower bounds correspond to incoherent states with no superposition, while the upper limits are reached by maximally coherent states, which are equal superpositions of all basis states. These limits help us interpret the coherence present in different quantum states and systems.
To illustrate the behavior of quantum coherence in our system, we plot the $l_1$-norm $\mathcal{C}_l(\rho)$ as a function of relevant physical parameters, including the magnetic field, temperature, and anisotropy. These plots allow us to directly visualize the persistence and variation of coherence under different conditions. All values of $\mathcal{C}_l$ are bounded according to equation~\eqref{Cl_bound}, with the lower limit corresponding to incoherent states and the upper limit to maximally coherent states.

\begin{figure*}
\centering
\subfloat[]{
\includegraphics[scale=0.36]{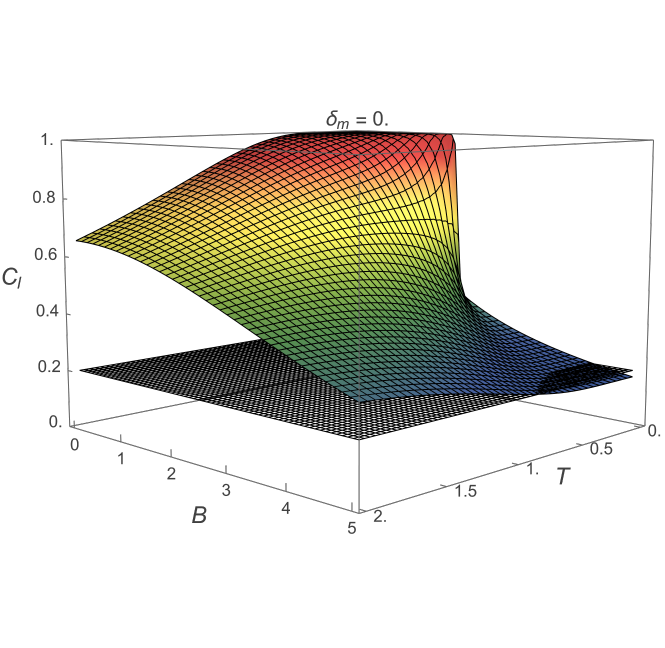}} 
\subfloat[]{
\includegraphics[scale=0.36]{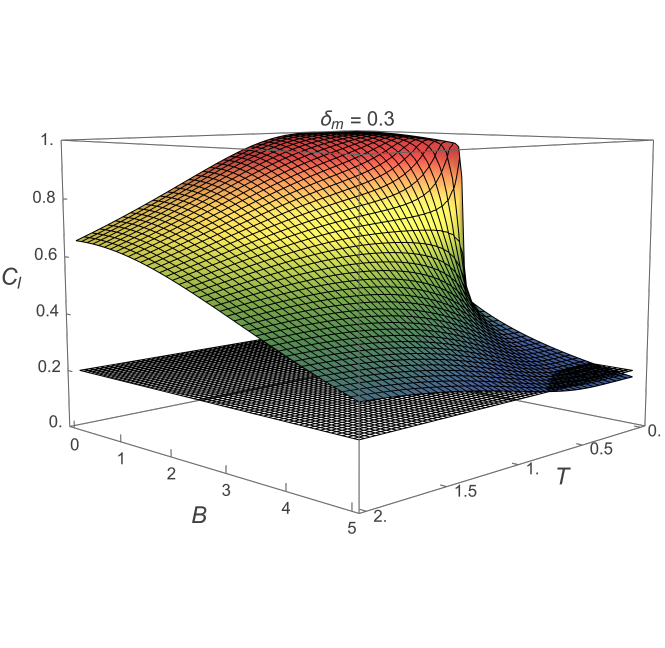}}
\subfloat[]{
\includegraphics[scale=0.36]	{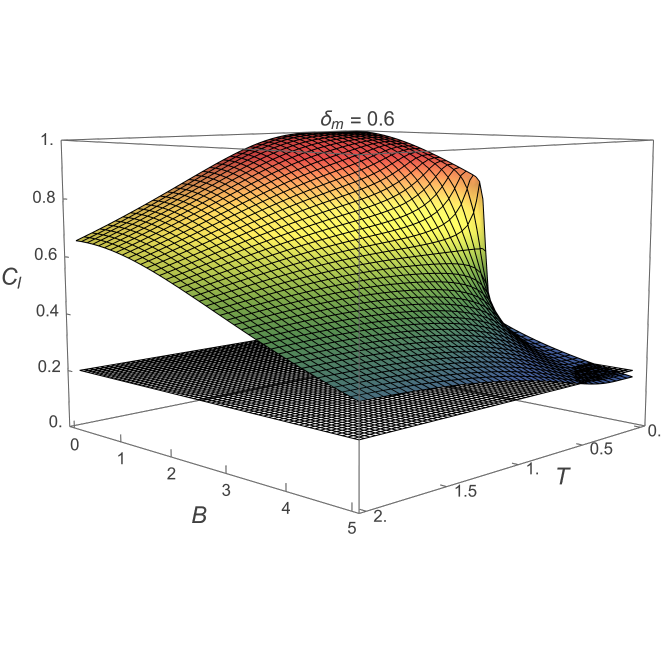}} 
\subfloat[]{
\includegraphics[scale=0.36]{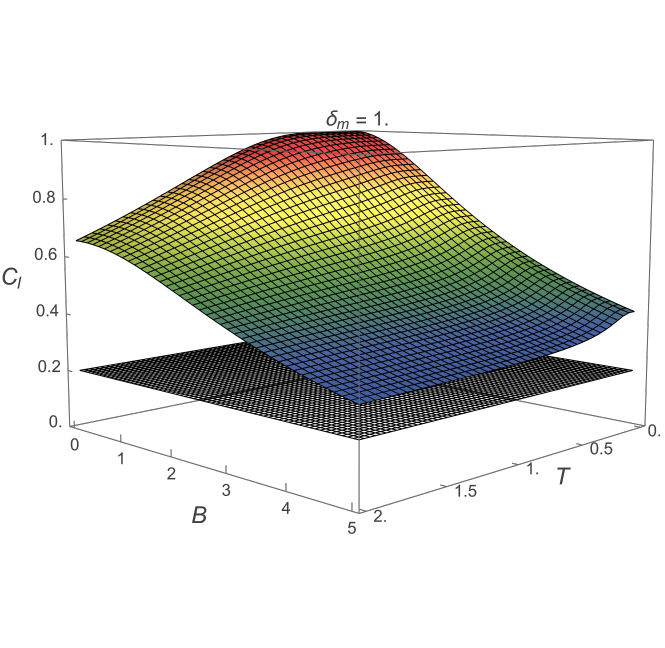}}
\caption{Quantum coherence $C_l$ as a function of temperature $T \in [0,2]$,
and magnetic field $B \in [0,5]$ for $\delta_c=0.9$, $J=1$, and $D=2$.
Panels (a)--(d) correspond to $\delta_m=0$, $0.3$, $0.6$, and $1$, respectively.}
%
\label{fig2} 
\end{figure*}

Figure \ref{fig2}  illustrates the behavior of quantum coherence $\mathcal{C}_l$ as a function of temperature $T$, and magnetic field $B$ for $\delta_c = 0.9$, $D = 2$, and four values of magnetic anisotropy ($\delta_m = 0, 0.3, 0.6, 1$). It is evident that increasing $\delta_m$ enhances coherence and makes it more robust against temperature-induced decay. In particular, when magnetic anisotropy is strong (e.g., $\delta_m = 1$), coherence persists over a broader range of $T$ and $B$. Conversely, with no magnetic anisotropy ($\delta_m = 0$), coherence decreases rapidly as temperature rises. These results highlight the crucial role of magnetic anisotropy in preserving quantum coherence, an essential resource for spin-based quantum technologies.
From a physical perspective, the enhancement of quantum coherence with increasing $\delta_m$ originates from the anisotropic spin–spin interactions, which favor the persistence of quantum superpositions. Magnetic anisotropy lifts degeneracies in the energy spectrum and strengthens the off-diagonal elements of the density matrix, which directly contribute to the $l_1$-norm of coherence. As a result, coherence becomes more resilient to thermal fluctuations. On the other hand, increasing temperature populates higher-energy eigenstates through thermal mixing, reducing the off-diagonal density-matrix elements and thus suppressing quantum coherence. This competition explains the observed robustness of coherence at strong anisotropy and its decay at elevated temperatures.

\begin{figure*}
\centering
\includegraphics[scale=0.25]{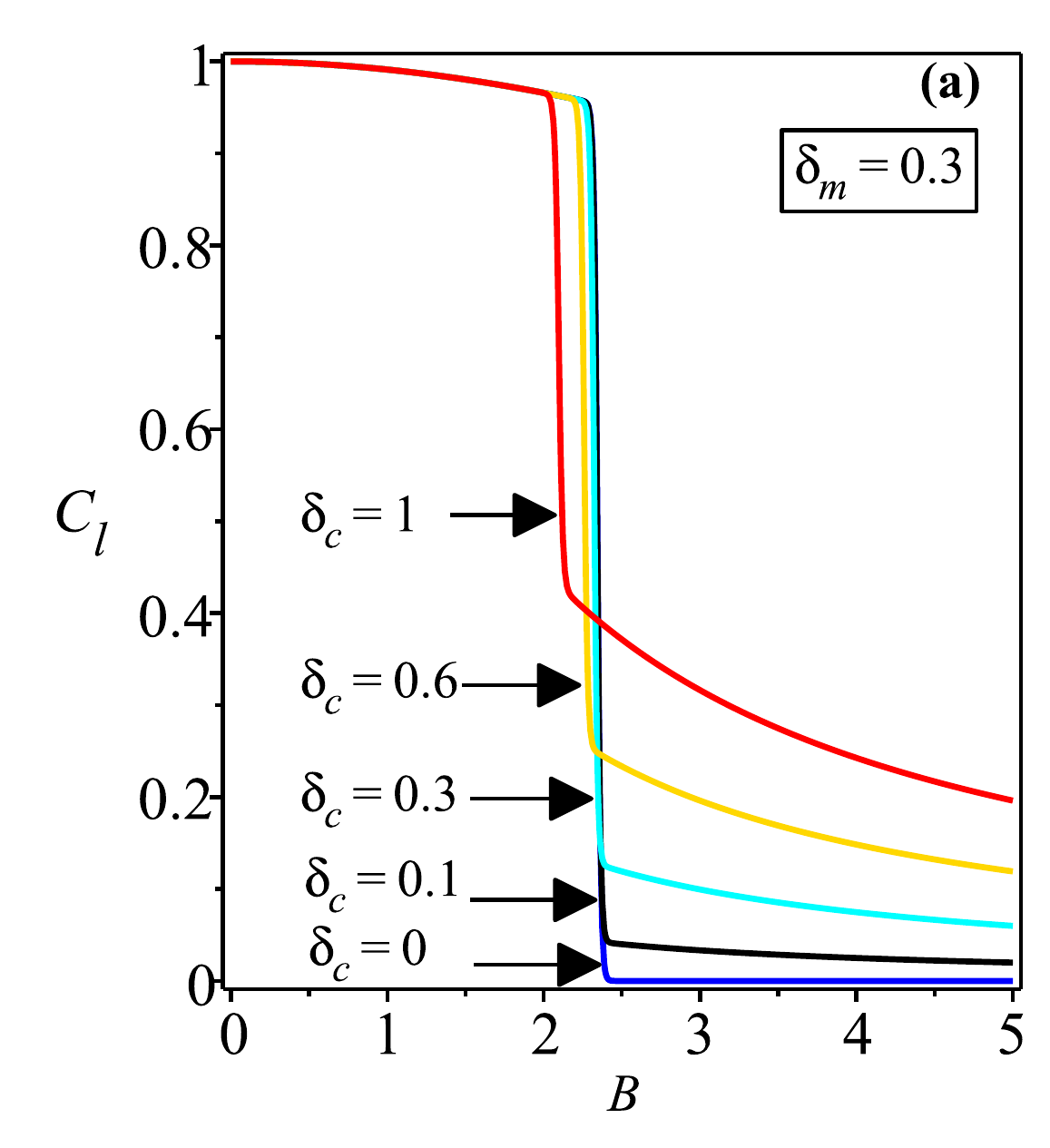} \quad
\includegraphics[scale=0.25]{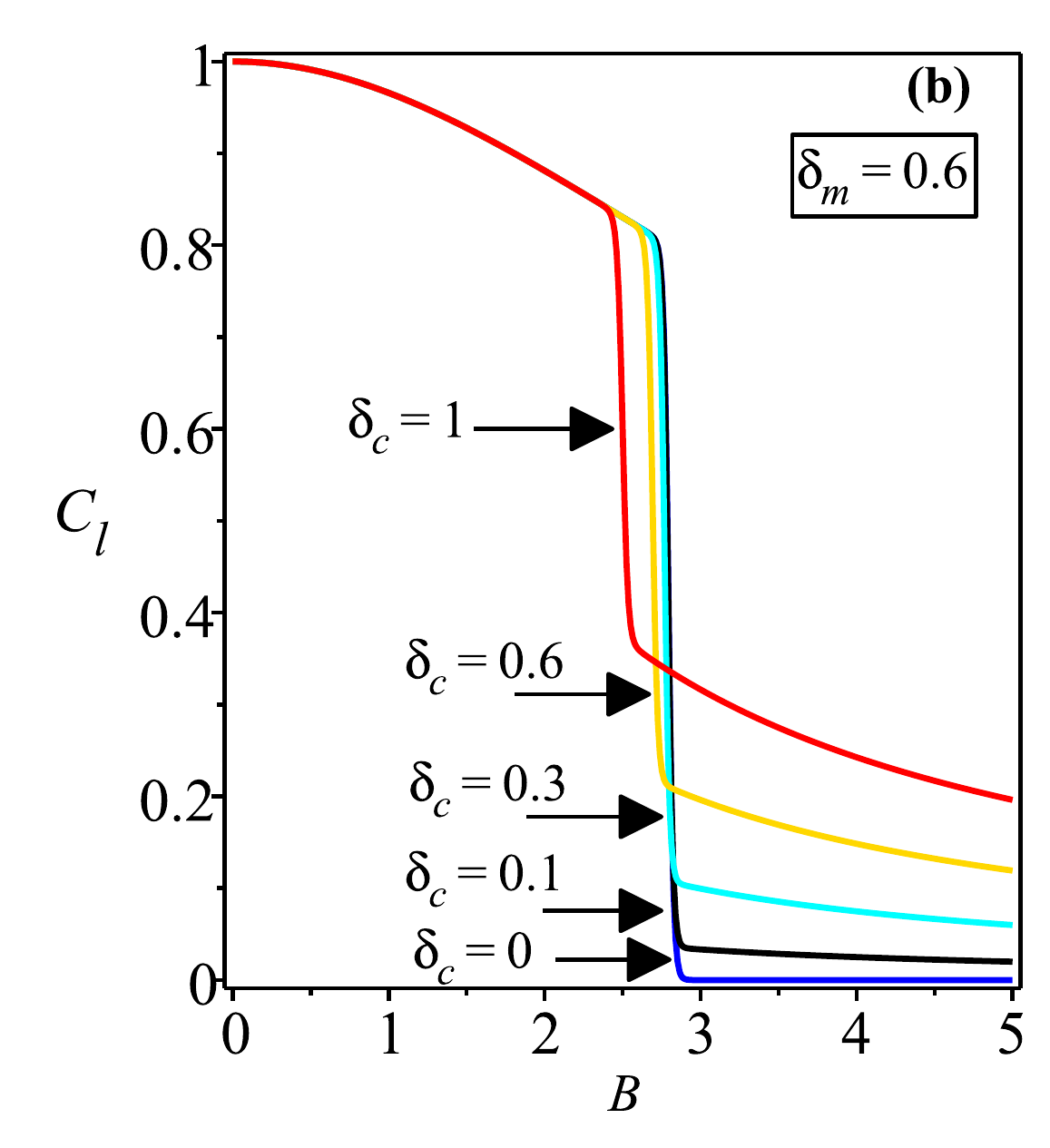}\quad
\includegraphics[scale=0.25]{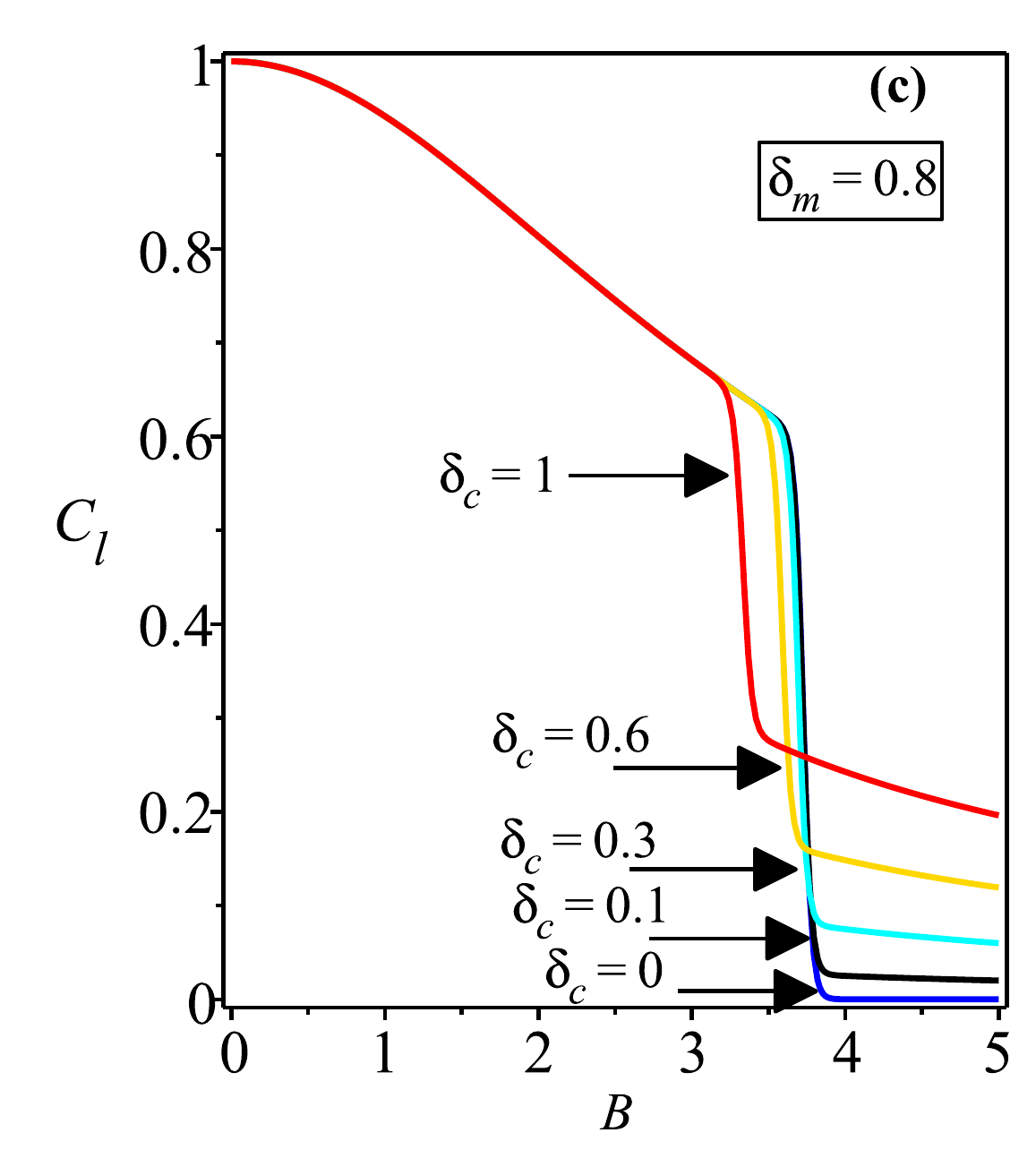} 
\caption{Effect of coupling anisotropy $\delta_c$ on the coherence $C_l$ as a function of the magnetic field $B$ for $D=2$, and $J=1$. Panels (a)–(c) correspond to $\delta_m = 0.3$, $0.6$, and $0.8$, respectively.}
\label{fig11} 	
\end{figure*}

Figure \ref{fig11}  illustrates how coupling anisotropy $\delta_c$ affects quantum coherence $\mathcal{C}_l$ as a function of the magnetic field $B$. The results reveal two important effects. First, increasing $\delta_c$ {promotes} the revival of coherence once the magnetic field exceeds a critical value $B_c$. Notably, $B_c$ decreases as $\delta_c$ increases, indicating that stronger coupling anisotropy allows coherence to recover at weaker magnetic fields. 
Second, magnetic anisotropy $\delta_m$ exhibits a different role. It shifts $B_c$ to higher values and generally enhances coherence. Overall, these findings highlight a {competition} between the two types of anisotropy. Coupling anisotropy $\delta_c$ promotes the recovery of coherence after the transition, while magnetic anisotropy $\delta_m$ stabilizes coherence and smooths the transition, providing a more robust quantum state.

\begin{figure*}
\centering
\includegraphics[scale=0.25]{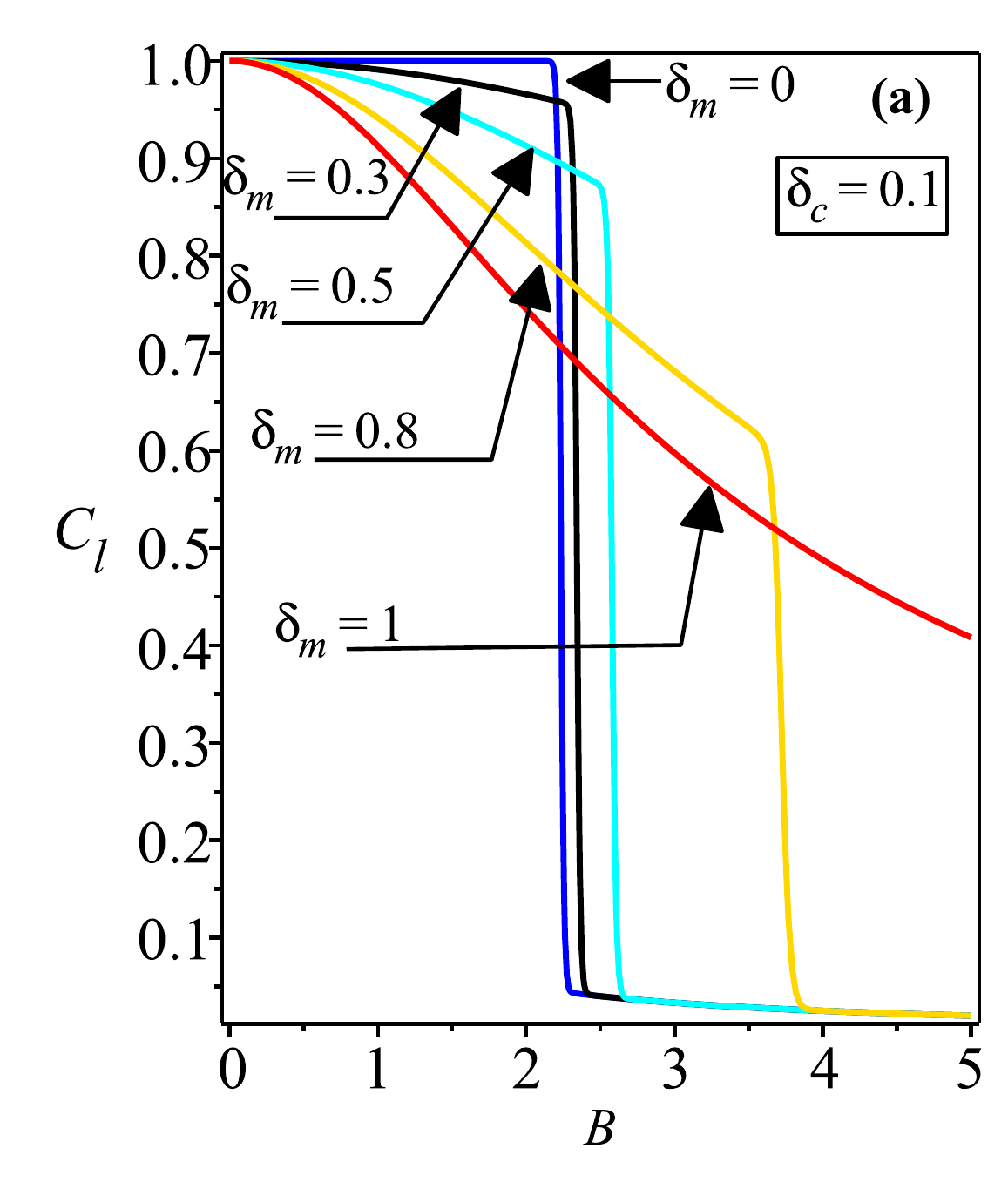} \quad
\includegraphics[scale=0.26]{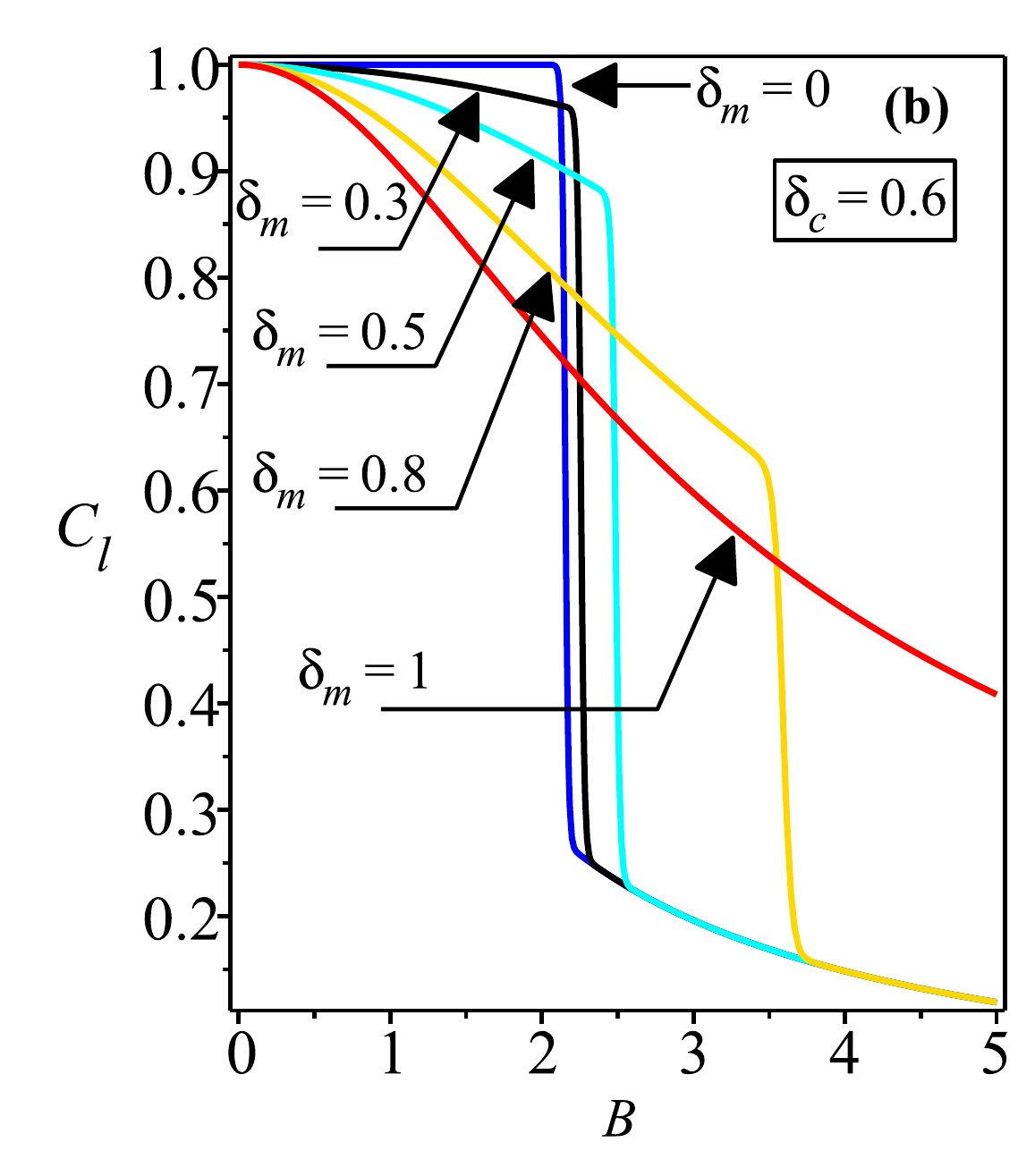} \quad
\includegraphics[scale=0.26]{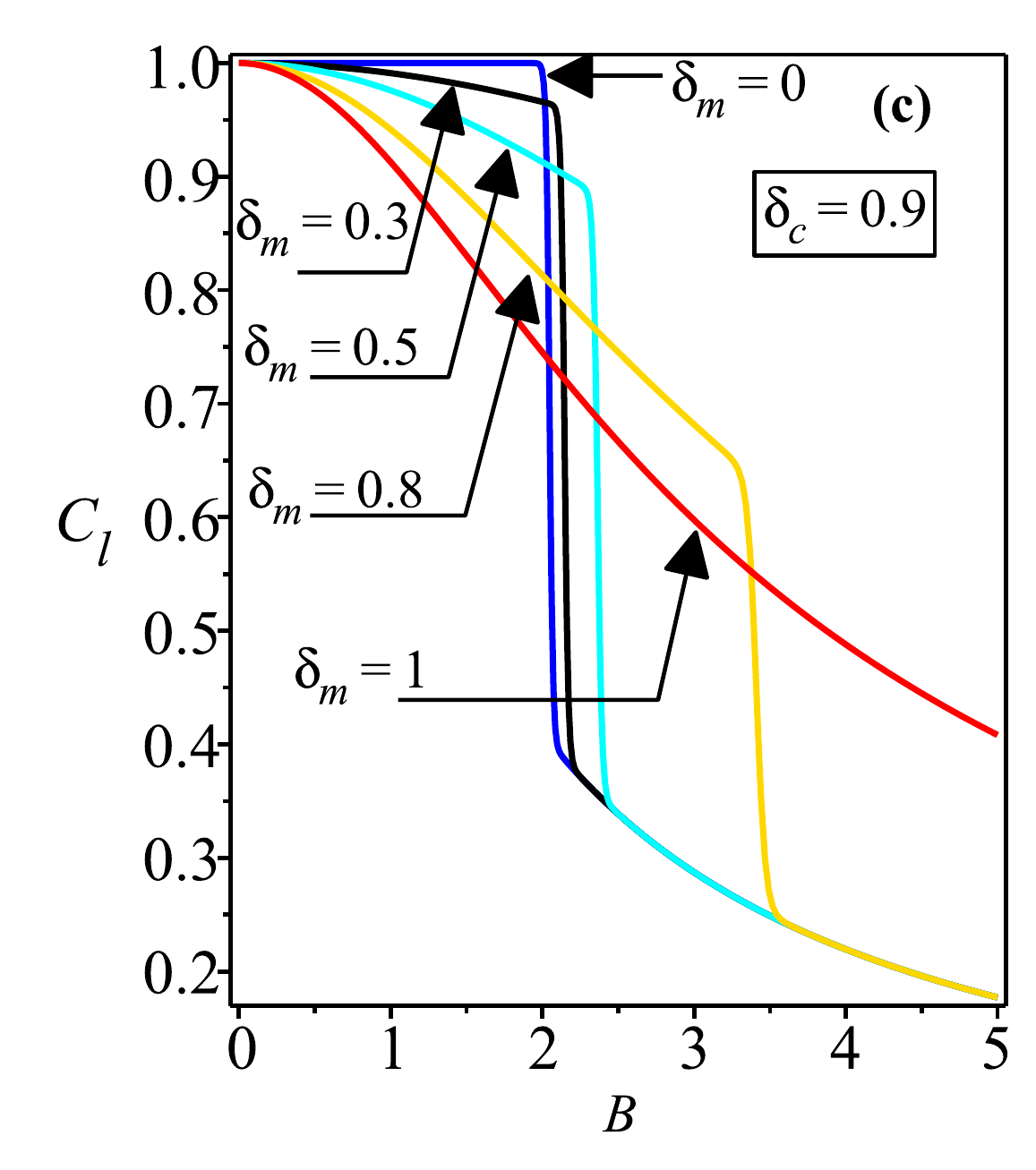} 
\caption{Effect of magnetic anisotropy $\delta_m$ on the coherence $C_l$ as a function of the magnetic field $B$ for $D=2$, and $J=1$. Panels (a)–(c) correspond to $\delta_c=0.1$, $0.6$, and $0.9$, respectively.}
\label{fig22}  
\end{figure*}

Figure \ref{fig22}  illustrates the effect of magnetic anisotropy $\delta_m$ on quantum coherence $\mathcal{C}_l$ as the magnetic field $B$ varies. A key observation is that increasing $\delta_m$ shifts the critical field $B_c$ to higher values while simultaneously enhancing coherence across all values of coupling anisotropy $\delta_c$.
{This effect becomes particularly pronounced for larger values of $\delta_c$.}
For instance, when $\delta_c = 0.9$, the revival of coherence becomes significantly stronger.
These results complement those shown in Figure \ref{fig11}, demonstrating that magnetic and coupling anisotropies can act together to stabilize quantum coherence. This combined effect provides an effective strategy for designing quantum systems with more robust ground-state coherence, a valuable resource for quantum information processing.

In summary, Figures \ref{fig2}–\ref{fig22} collectively demonstrate that magnetic and coupling anisotropies play complementary roles in controlling quantum coherence $\mathcal{C}_l$. Magnetic anisotropy $\delta_m$ enhances coherence and shifts critical fields, making coherence more robust against thermal and field-induced fluctuations, while coupling anisotropy $\delta_c$ facilitates the revival of coherence after transitions at weaker magnetic fields. {The interplay between these two forms of anisotropy therefore provides an effective mechanism for stabilizing and manipulating coherence in spin-based quantum systems.}
We emphasize that this enhancement is valid within the parameter regime where the system remains in thermal equilibrium and is well described by the reduced density matrix formalism. At very high temperatures or extremely strong magnetic fields, thermal mixing dominates and coherence is inevitably suppressed. Therefore, anisotropy-induced coherence stabilization is most effective at low-to-intermediate temperatures, {which is the regime most relevant for} solid-state spin systems. These findings underscore the potential of anisotropy engineering as a practical strategy for preserving quantum coherence in realistic experimental conditions.

\section{Hierarchy of quantum correlations and coherence \label{sec5}}

We numerically explore the hierarchy between different quantum correlations and coherence measures in our system. {We first consider} the case where the magnetic field is absent ($B = 0$). Under this condition, the density matrix elements simplify to: $\rho_{11} = \rho_{44} = \cosh(E_3/T)$, $\rho_{22} = \rho_{33} = \cosh(E_1/T)$, $|\rho_{14}| = \sinh(E_3/T)$, and $|\rho_{23}| = \sinh(E_1/T)$. These expressions allow us to compute various quantum indicators in terms of temperature $T$ and energy levels $E_1$ and $E_3$. The key quantities—concurrence, $l_1$-norm coherence, Bell-CHSH observable $\mathbb{B}$, and local quantum uncertainty (LQU) are given by
\begin{align}
&	C[\rho](T)=\frac{2}{\mathcal{Z}}\max[0,|\rho_{23}|-\rho_{11},|\rho_{14}|-\rho_{22}],\label{633}\\
&	\mathcal{C}_{l_1}[\rho](T)= \frac{2}{\mathcal{Z}}(|\rho_{23}|+|\rho_{14}|),\label{644}\\
&	\mathbb{B}[\rho](T)=\frac{4\sqrt{2}}{\mathcal{Z}}\sqrt{|\rho_{14}|^2+|\rho_{23}|^{2}},\label{655}\\
&	\text{LQU}[\rho](T)= 1-\frac{3+2\rho_ {11}\rho_{22}+2|\rho_{14}\rho_{23}|}{\mathcal{Z}\sqrt{3/2+\rho_{11}\rho_{22}}}.\label{666}
\end{align}
These expressions clearly show how to examine the thermal behavior of various quantum features. {In particular, they show how temperature influences the interplay between coherence and entanglement, as well as the behavior of these quantities in the absence of an external magnetic field.} By comparing these measures, we can identify which quantum resource is more resistant to thermal fluctuations, allowing us to better understand their interactions in practical quantum environments.

\begin{figure*}
\centering
\includegraphics[scale=0.25]{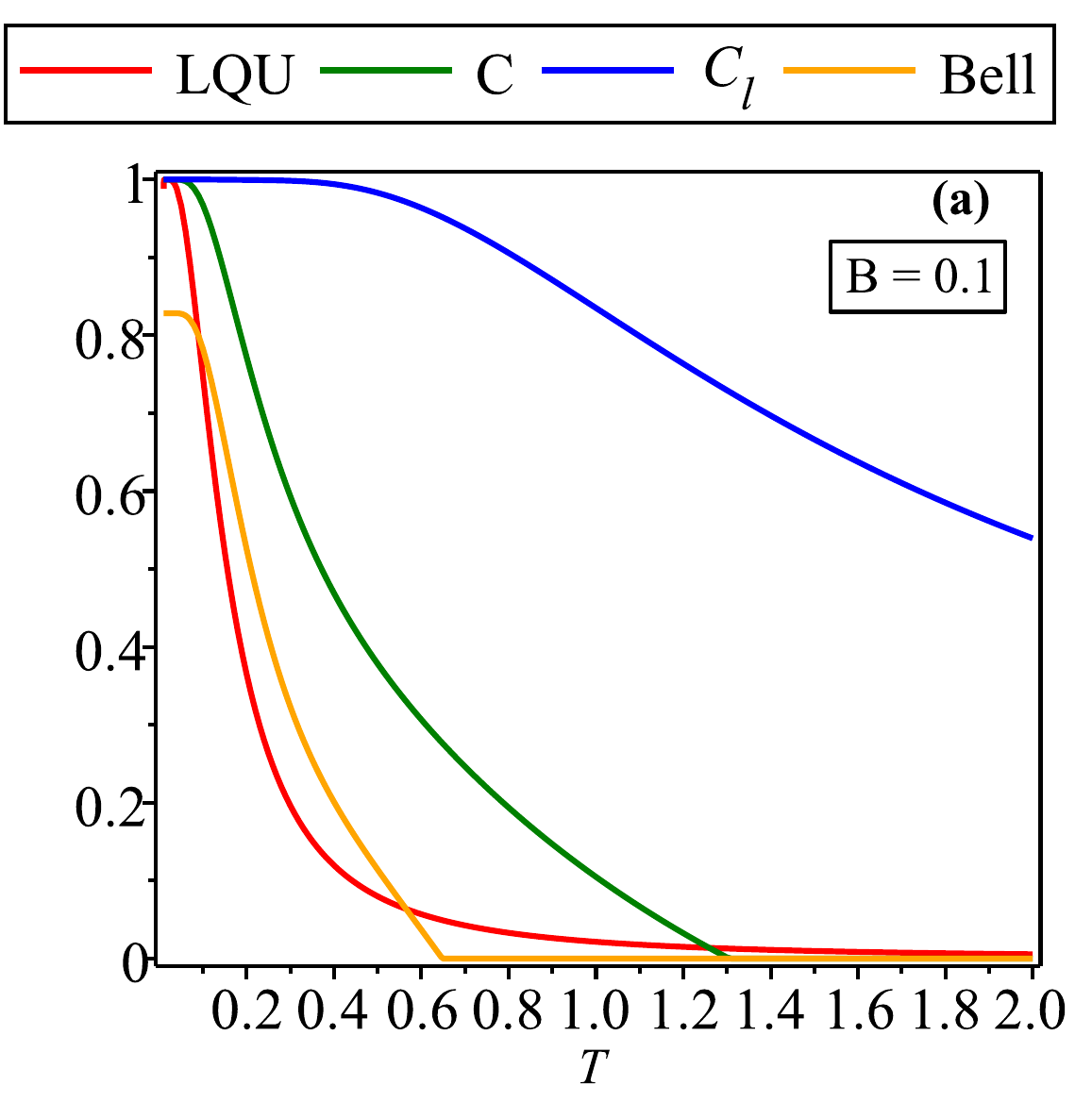} \quad
\includegraphics[scale=0.24]{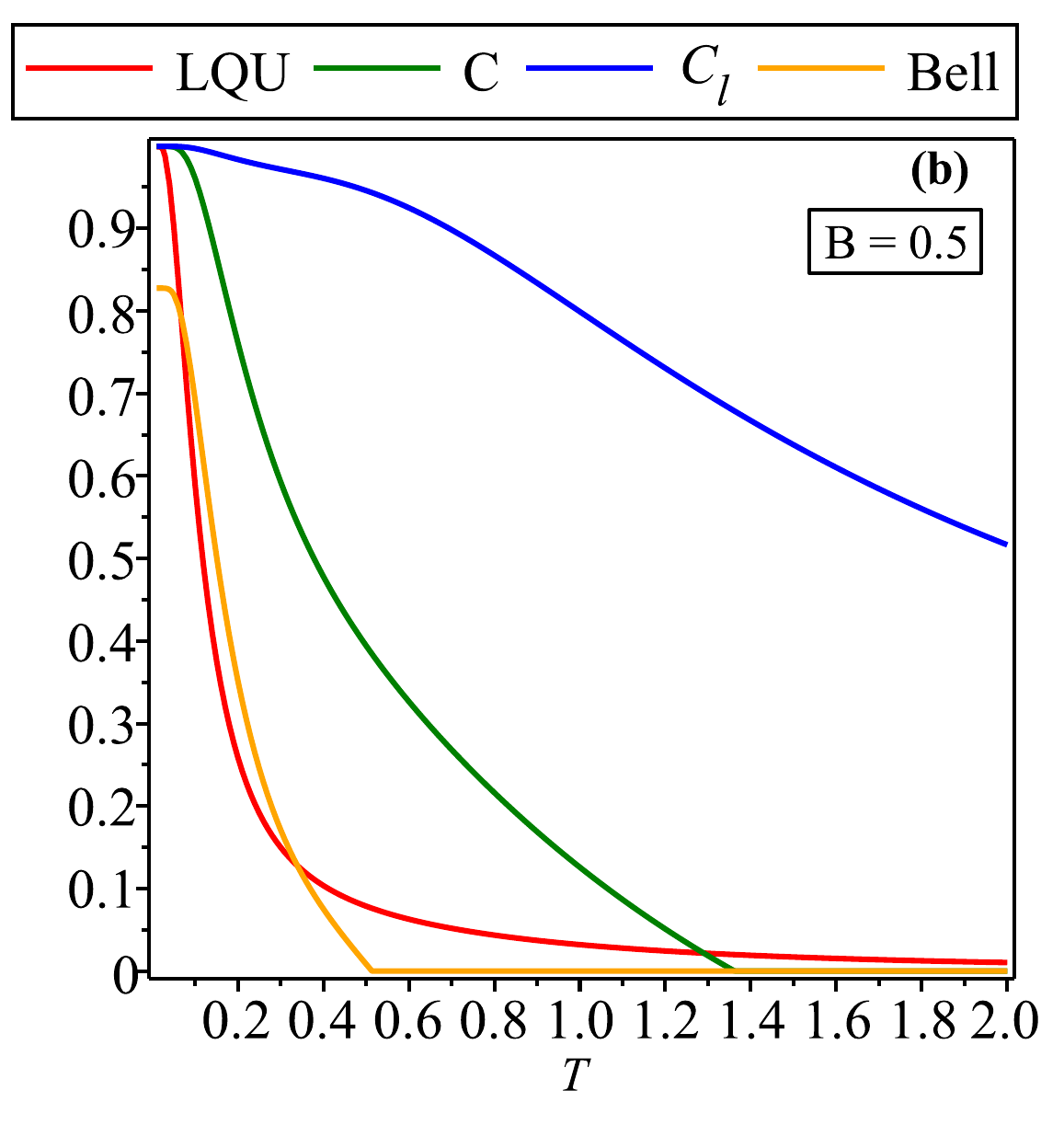} \quad
\includegraphics[scale=0.246]{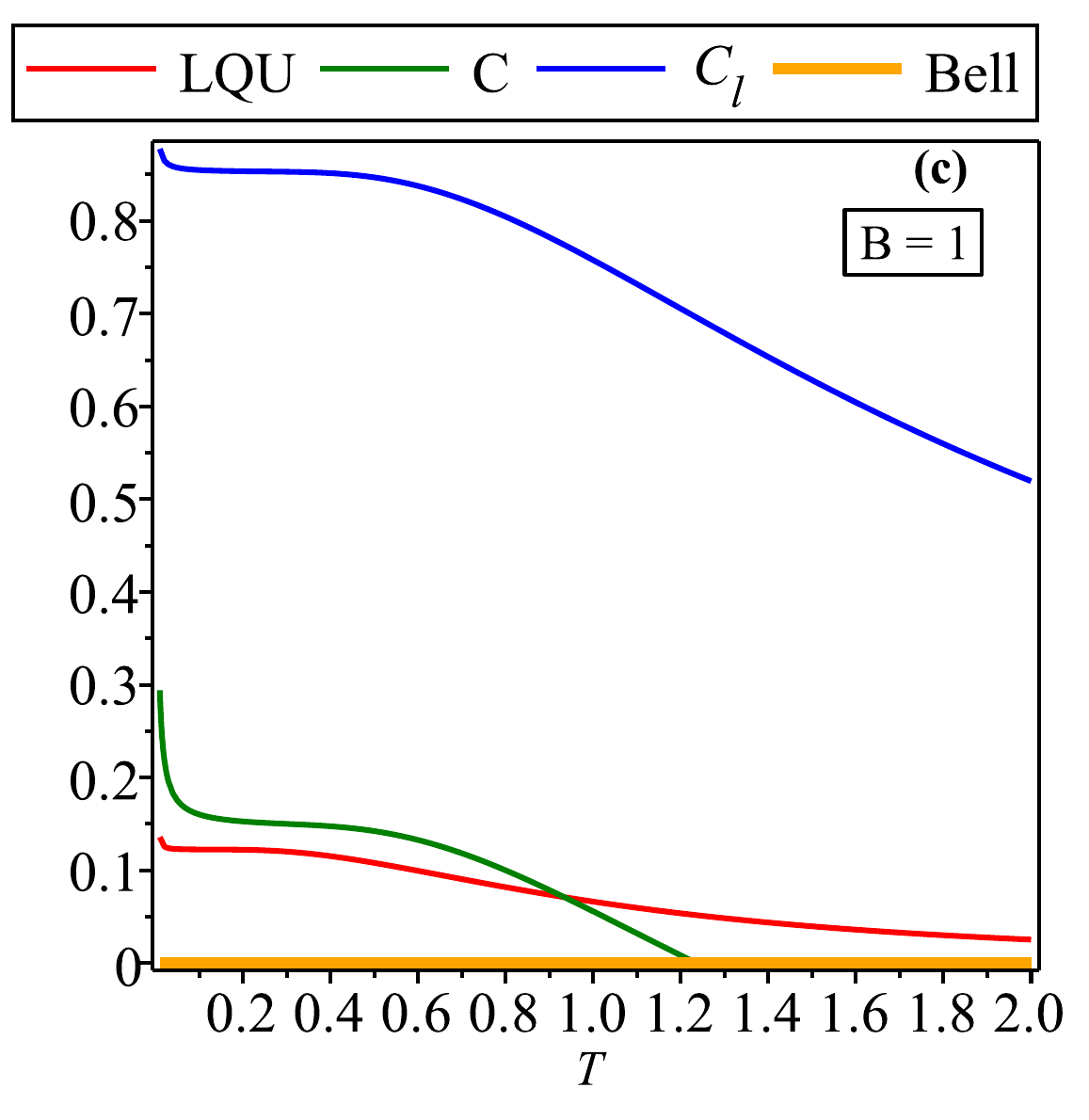} 
\caption{Effect of the magnetic field $B$ on the hierarchy of quantum correlations, and coherence as a function of temperature $T$ for $D = 1$, $J = 1$, $\delta_m = 0.1$, and $\delta_c = 1$. Red line: local quantum uncertainty (LQU)  (Eq.~\eqref{633}), green line: concurrence $C$ (Eq.~\eqref{644}), blue line: quantum coherence $\mathcal{C}_l$ (Eq.~\eqref{655}), orange line: Bell–CHSH observable $\mathbb{B}$ (Eq.~\eqref{666}).}
%
\label{fig4} 
\end{figure*}

Figure~\ref{fig4} presents a systematic study of quantum correlations and coherence. It displays the local quantum uncertainty (LQU), concurrence $C$, coherence $C_{l}$, and Bell-CHSH observable $\mathbb{B}$ versus the temperature $T$ for three values of magnetic field $B$. The analysis reveals a clear hierarchy in the thermal degradation of quantum properties. {For $B=0.1$, three critical temperatures can be identified}. (1): $T_{c1}=0.65$, where $\mathbb{B}$ vanishes, indicating the loss of Bell nonlocality. (2): $T_{c2}=1.3$, where $C$ drops to zero, marking the disappearance of entanglement. (3): $T_{c3}=1.8$, where LQU reaches zero, signifying the complete loss of local quantum correlations. In particular, coherence $C_{l}$ shows a more gradual decay, starting at $T=0.48$, and persisting beyond the other quantum measures. For $T<T_{c1}$, the system maintains maximum values of all quantum measures, confirming a strongly correlated quantum state. When $B$ increases, this degradation pattern persists but with modified critical temperatures, and generally reduced initial correlation strengths. These results show that the coherence $C_{l}$ is the most thermally robust quantum resource, followed by {local quantum correlations}, entanglement, and finally Bell nonlocality. These findings have important implications for the design of quantum devices, in particular for understanding how temperature and magnetic fields affect different quantum resources in anisotropic systems. In conclusion, this hierarchy $\mathbb{B} \rightarrow C \rightarrow \text{LQU}$ in thermal degradation provides a fundamental framework for predicting the behavior of quantum systems in realistic environments.

\begin{figure*}
\centering
\includegraphics[scale=0.25]{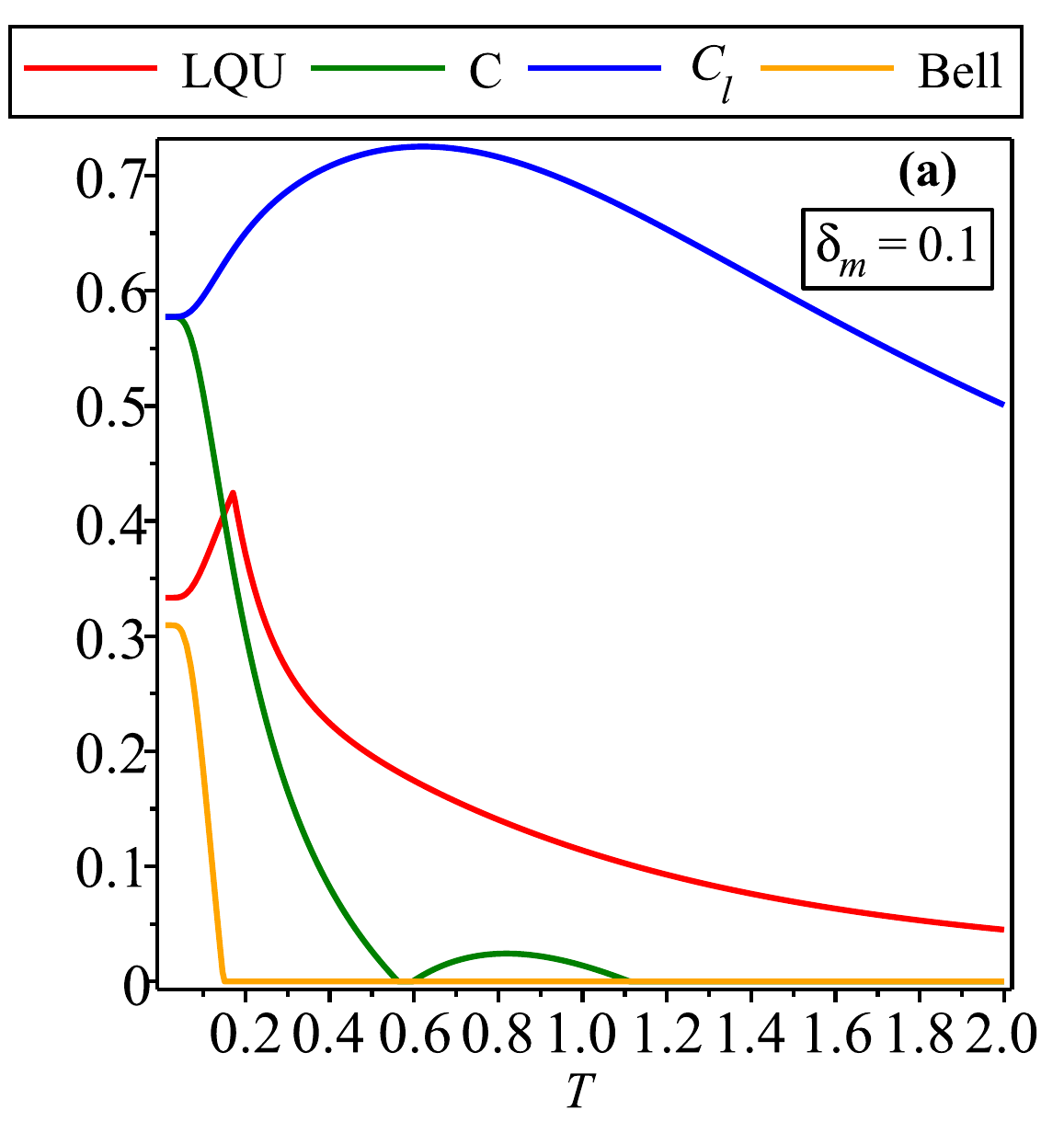} \quad
\includegraphics[scale=0.267]{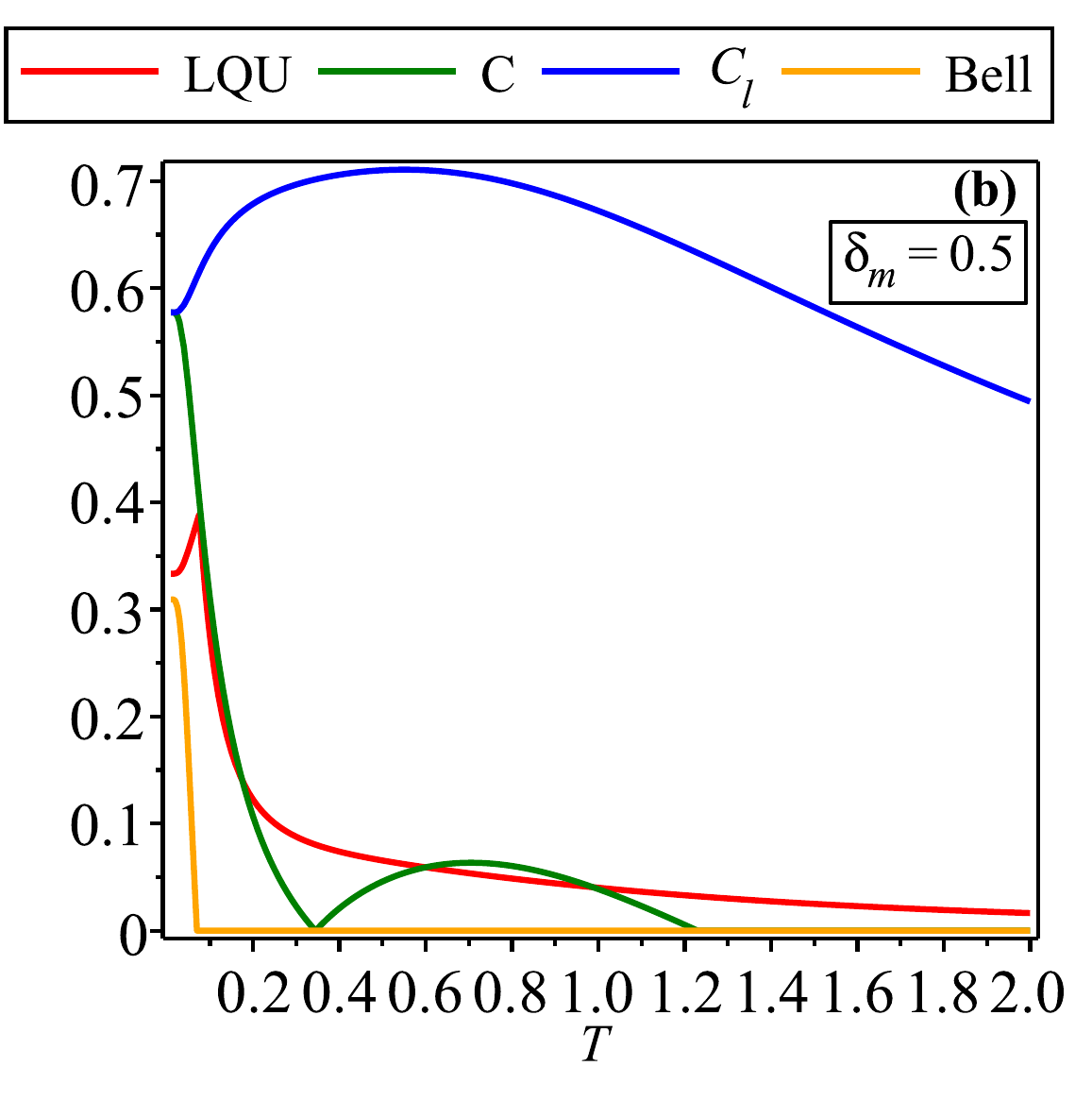} \quad
\includegraphics[scale=0.277]{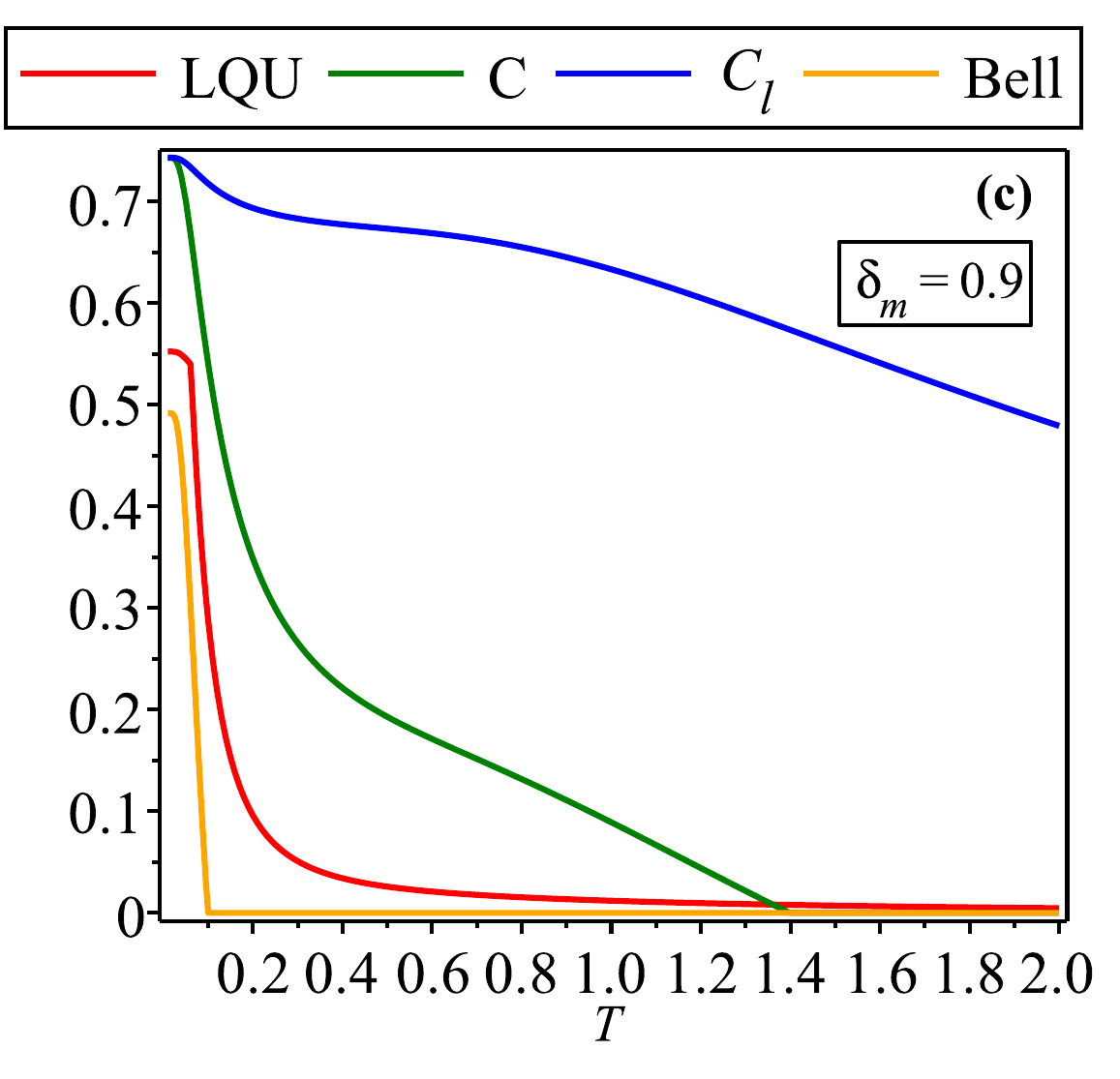} 
\caption{Effect of magnetic anisotropy $\delta_m$ on the hierarchy of quantum correlations, and coherence for $D = 1$, $J = 1$, $B = \sqrt{2}$, and $\delta_c = 1$. Red line: local quantum uncertainty (LQU)  (Eq.~\eqref{633}), green line: concurrence $C$ (Eq.~\eqref{644}), blue line: quantum coherence $\mathcal{C}_l$ (Eq.~\eqref{655}), orange line: Bell–CHSH observable $\mathbb{B}$ (Eq.~\eqref{666}).}
%
\label{fig5} 
\end{figure*}

Figure~\ref{fig5} examines the influence of magnetic anisotropy \(\delta_m\) on quantum correlations and coherence by plotting their quantifiers-concurrence \(C\), local quantum uncertainty (LQU), Bell-CHSH observable $\mathbb{B}$, and coherence \(C_l\)) versus the temperature (\(T\)) for \(D=1\), \(J=1\), \(B=\sqrt{2}\), and  \(\delta_c=1\). For \(\delta_m=0.1\), the system exhibits two critical temperatures: \(T_{c1}\), where \(C\) first increases to a peak before decaying to zero at \(T_{c2}\). {Meanwhile}, LQU stabilizes at a non-zero value, indicating residual quantum correlations. Notably, \(\mathbb{B}\) remains zero for all \(T\), confirming separability, while \(C_l\) shows robust coherence. As \(\delta_m\) increases to 0.9, entanglement (\(C\)) and coherence (\(C_l\)) are significantly enhanced, although \(LQU\) improves only at low temperatures before decaying. These results demonstrate that higher magnetic anisotropy (\(\delta_m\)) not only {enhances} entanglement and coherence but also modulates the hierarchy of quantum correlations, with \(C_l\) emerging as the most resilient resource. The results highlight \(\delta_m\) as a critical control parameter for optimizing quantum resources in thermally fluctuating environments.

\begin{figure*}
\centering
\includegraphics[scale=0.25]{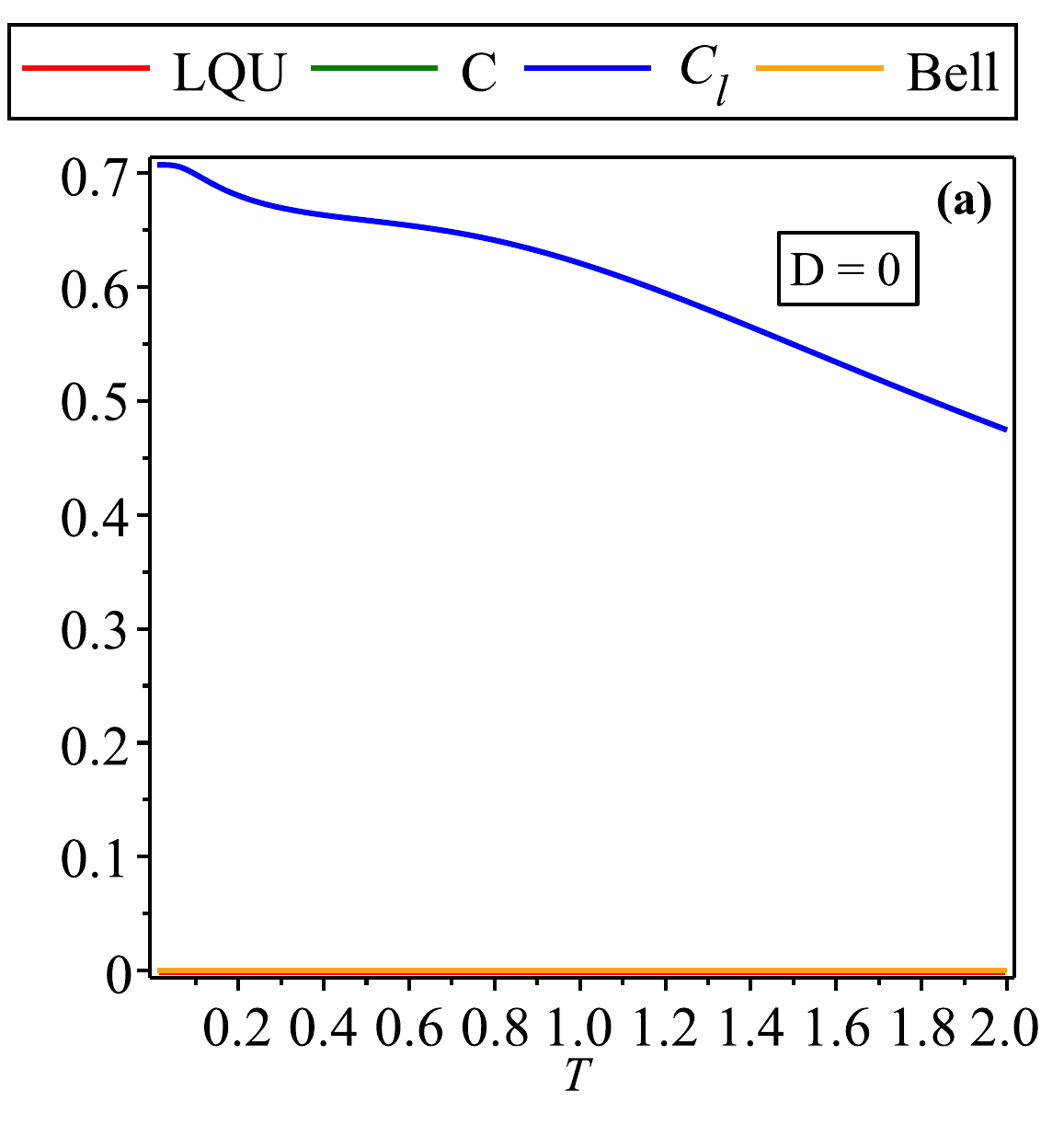}\quad
\includegraphics[scale=0.25]{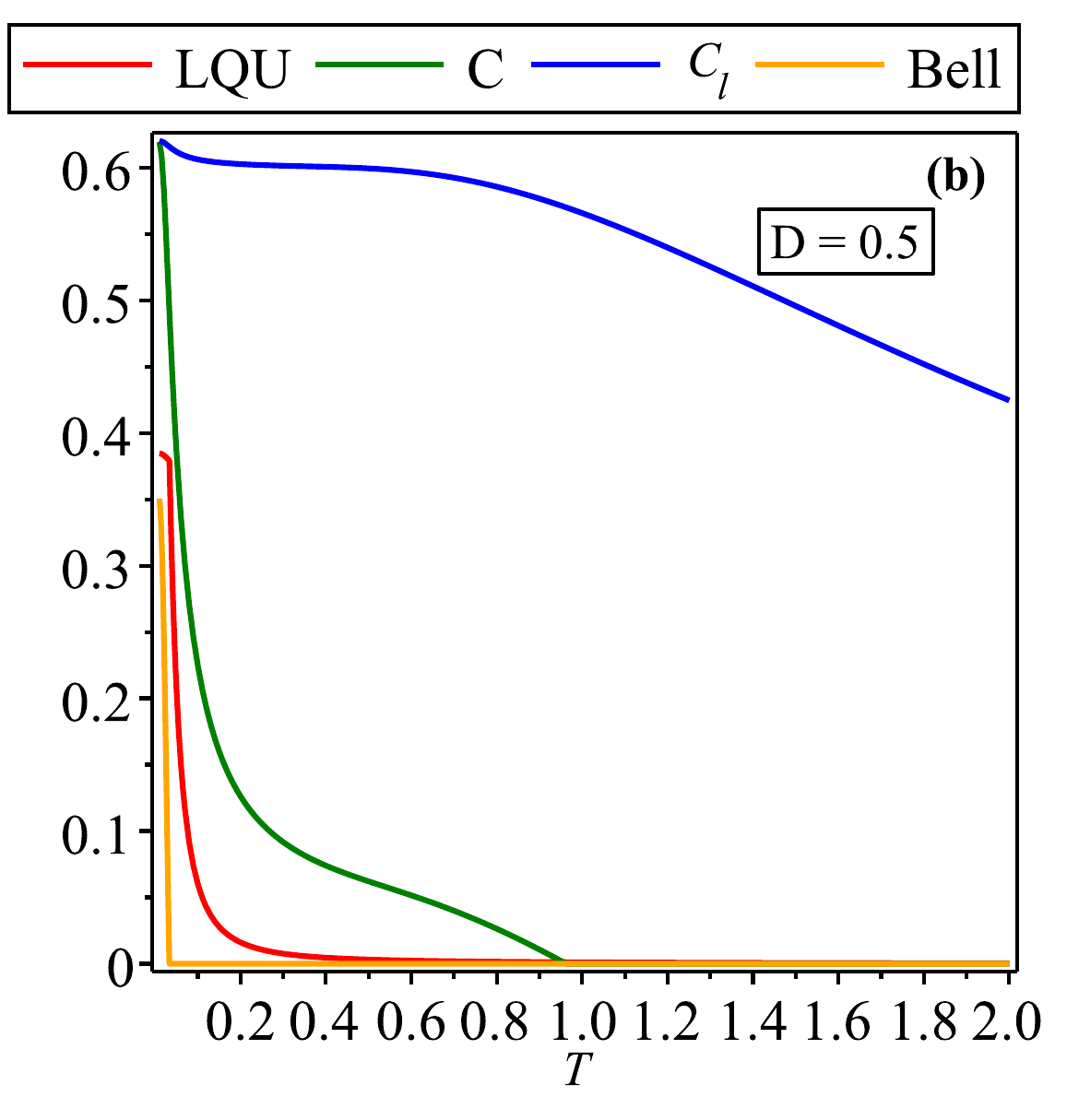} \quad
\includegraphics[scale=0.25]{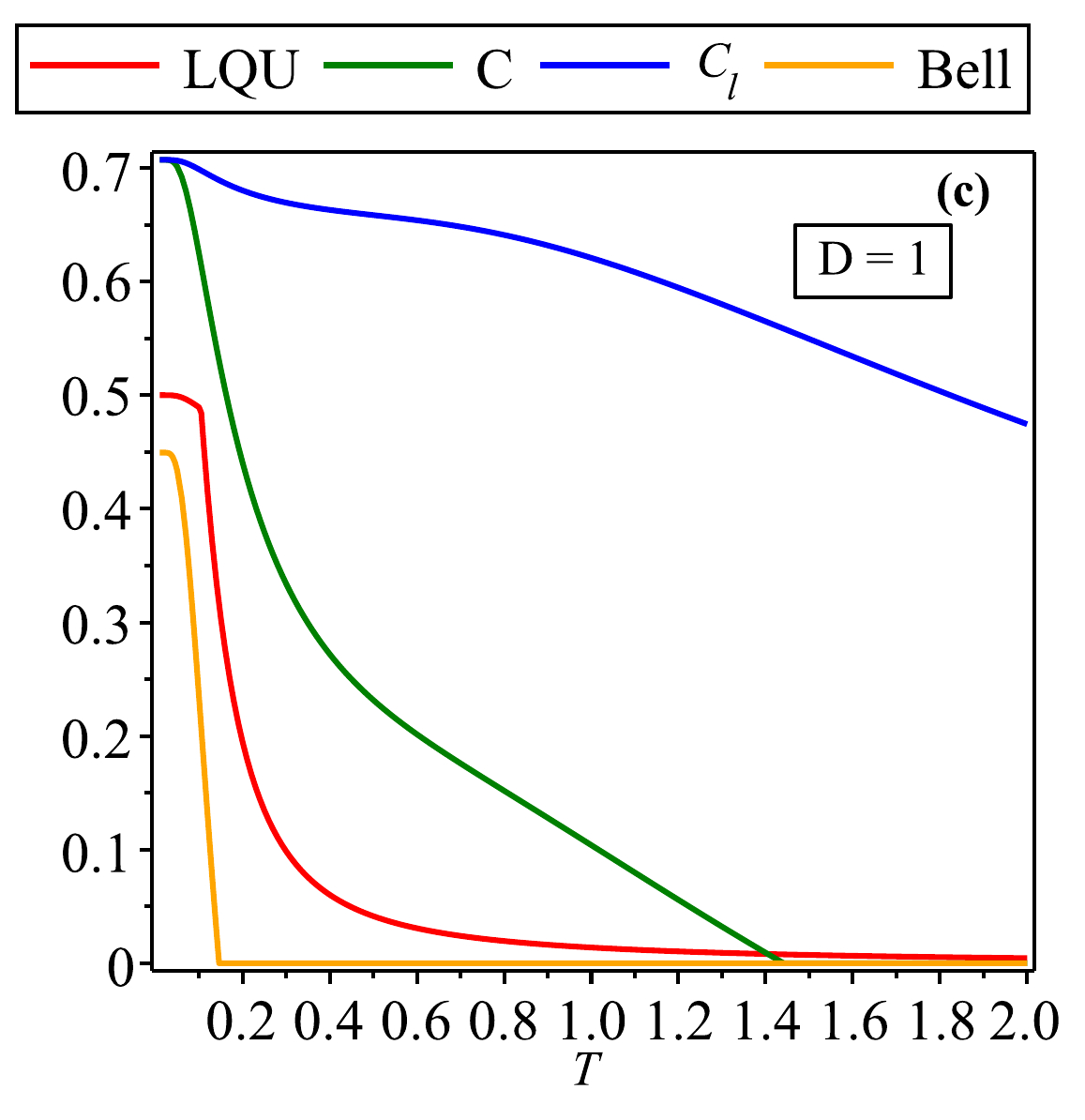}
\caption{Effect of the Dzyaloshinskii–Moriya (DM) interaction $D$ on the hierarchy of quantum correlations, and coherence for $B = \sqrt{2}$, $J = 1$, $\delta_m = 1$, and $\delta_c = 1$. Red line: local quantum uncertainty (LQU) (Eq.~\eqref{633}), green line: concurrence $C$ (Eq.~\eqref{644}), blue line: quantum coherence $\mathcal{C}_l$ (Eq.~\eqref{655}), orange line: Bell–CHSH observable $\mathcal{B}$ (Eq.~\eqref{666}).}
%
%
\label{fig6} 
\end{figure*}

Figure~\ref{fig6} explores the role of the Dzyaloshinskii-Moriya (DM) interaction (\(D\)) in modulating quantum correlations, and coherence within a two-qubit XY Heisenberg model, considering the effects of magnetic anisotropy \(\delta_m\), and coupling anisotropy \(\delta_c\). When \(D = 0\), the system remains separable (unentangled) at all temperatures, highlighting the necessity of the DM interaction for  generating quantum correlations. As \(D\) increases, quantum correlations are {restored}, and persist up to a critical temperature \(T_c\) before undergoing a gradual thermal decay. The maximum correlation strength occurs for \(\delta_m = 1\), \(D = 1\), and \(T = 0\), demonstrating the synergistic enhancement of quantum properties by DM interaction and anisotropy. Beyond \(T_c\), the correlations decrease and eventually disappear at higher temperatures. These results highlight the dual importance of DM interaction and anisotropy in sustaining quantum correlations, with the former enabling their emergence and the latter enhancing their robustness. {They also provide useful insight into the design of quantum systems in which tunable interactions, such as \(D\), and anisotropy can be exploited to preserve correlations in thermal environments.}

\section{Comparison with Previous Studies}\label{V}

Comparing our results with the established coherence and quantum correlation relations from existing research yields useful results. The study in~\cite{PRA2017} established a unified framework that explains quantum coherence through its connection to discord-like correlations and entanglement. Coherence is a broader representation of quantumness, while entanglement is a limited expression of nonclassical correlations. Discord-like correlations occupy an intermediate position since they can persist even after entanglement disappears.
%
Our results for the thermal Heisenberg $XY$  model demonstrate complete agreement with the general framework. Coherence shows greater resilience because the $l_1$-norm coherence remains present across a wide range of temperatures and magnetic fields, exceeding the operational range of concurrence. Local quantum uncertainty (LQU), which detects discord-like quantum correlations, exists across parameter ranges where entanglement declines and where Bell-CHSH violations become undetectable. This study shows that different quantumness indicators exhibit different levels of response to thermal effects because they tend to vanish at distinct thermal points.
%
As the temperature rises, Bell nonlocality is the first quantum characteristic to disappear, followed by the loss of entanglement. The system maintains quantum correlations at local distances until it reaches the extreme of its parameter range. However, coherence endures the longest, making it the most effective defense against thermal mixing. This established sequence confirms that Bell nonlocality is the strongest and most delicate indicator of quantum behavior. Entanglement offers less strict limitations, but remains sensitive to decoherence. Coherence functions as the fundamental quantum resource that endures when all stronger quantum connections have disappeared.
%
Another important point is that this hierarchy can be observed not only qualitatively but also quantitatively in our model through the dependence of different measures on temperature, magnetic field, and anisotropy parameters. In this sense, the present system provides a concrete thermal realization of the abstract relations discussed in~\cite{PRA2017}. It clearly shows how coherence, entanglement, and discord-like correlations respond differently to thermal effects and external control parameters.
Therefore, our model offers an explicit example in which the general hierarchy among Bell nonlocality, entanglement, discord-like correlations, and coherence can be directly observed in a thermally driven spin system. This comparison further highlights the physical relevance of coherence as the most persistent quantum resource. It also supports the broader interpretation that nonclassical correlations beyond entanglement play an essential role in the finite-temperature behavior of low-dimensional quantum systems.

It is also informative to place our results in the broader context of the hierarchy of relations established for bipartite quantum correlations. Previous studies have shown that Bell nonlocality is a stronger, more protective form of quantum correlation than Einstein-Podolsky-Rosen (EPR) steering, which is stronger than entanglement \cite{Wiseman2007,Jones2007}. While every Bell nonlocal state is steerable and every steerable state is entangled, the reverse is not true in most situations. The various manifestations of nonclassicality create a well-established hierarchy among these forms of nonclassical behavior.
In addition, quantitative approaches to steering in two-qubit systems have been developed \cite{CostaAngelo2016}, making it possible to analyze more precisely the intermediate role played by steering between entanglement and Bell nonlocality. More recently, it has also been shown that coherence-based nonclassical correlations, such as the nonlocal advantage of quantum coherence, may persist beyond the range where Bell nonlocality is observed \cite{Mondal2018}. These results suggest that coherence-based signatures can provide a broader characterization of quantumness, complementing the information obtained from entanglement and Bell inequalities.
From this perspective, our results for the thermal Heisenberg $XY$ model follow the same qualitative hierarchy. In particular, Bell-CHSH violations disappear first as temperature increases, showing that nonlocality is the most fragile quantum feature in the present system. Entanglement, quantified by the concurrence, survives over a wider range of temperatures and control parameters than Bell nonlocality. Furthermore, the local quantum uncertainty (LQU), which measures discord-like quantum correlations, maintains its existence in all parameter regions where entanglement has disappeared. The $l_1$-norm coherence stands as the strongest measurement because it maintains its presence throughout the widest distribution of temperature and magnetic field conditions.
%
This ordering clearly shows that different quantum indicators do not disappear under thermal effects simultaneously, but instead follow a specific order of robustness. Bell nonlocality is lost first, followed by entanglement suppression. Discord-like correlations, and especially coherence, remain detectable for longer. This behavior aligns with the general theoretical framework, which states that nonlocality is the most restrictive signature of quantumness, while more general forms of quantum correlations can survive in wider parameter regimes.
Although steering is not explicitly evaluated in this work, the hierarchy observed in our results is consistent with the broader relations reported in the referenced literature. In this sense, our thermal two-spin model provides a concrete, physically transparent example of how different forms of quantum correlations and coherence are affected by temperature. The model also demonstrates that coherence and discord-like correlations can persist as valuable quantum resources even after stronger forms of nonclassicality, such as entanglement and Bell nonlocality, disappear.

\section{Conclusion}

{We have carried out a detailed investigation of} the impact of magnetic anisotropy $\delta_m$ on quantum correlations and coherence in the thermal equilibrium state of the Heisenberg XY model under the influence of a magnetic field $B$. We also examined the interplay of coupling anisotropy $\delta_c$ and the Dzyaloshinskii–Moriya (DM) interaction $D$. To obtain a comprehensive understanding of quantum behavior, multiple measures were employed. Indeed, concurrence $C$ was utilized to quantify entanglement, while local quantum uncertainty (LQU) was employed to measure local quantum correlations. Additionally, the Bell-CHSH observable $\mathbb{B}$ was considered to detect nonlocality, and the $l_1$-norm was employed to quantify quantum coherence $C_l$.
The results of our study indicate that magnetic anisotropy $\delta_m$ plays a central role in enhancing all quantum measures—$C$, $LQU$, $B$, and especially $C_l$—across a wide range of temperatures and interaction strengths.  {They also reveal a clear thermal hierarchy among these quantifiers.} Quantum coherence ($C_l$) exhibits the highest robustness, followed by entanglement ($C$). Local and nonlocal quantum correlations (LQU and $\mathbb{B}$) degrade more quickly with temperature. 
{This behavior is consistently observed across the different parameter regimes explored in this work and can be summarized by the hierarchy}
$
C_l > C > \mathrm{LQU} \approx \mathbb{B}.
$.


Moreover, the DM interaction $D$ was found to be indispensable for generating quantum correlations in this model. When $D = 0$, the system remains entirely separable, regardless of temperature. Once introduced, the DM interaction not only restores quantum correlations but also extends their persistence against thermal effects. Interestingly, we found that $\delta_m$ and $\delta_c$ do not act independently, rather, they exhibit a synergistic effect that strengthens and stabilizes coherence and correlations. For instance, increasing $\delta_c$ enhances the revival of coherence after a phase transition, while $\delta_m$ boosts robustness and shifts critical temperatures, effectively smoothing the transition itself.
Finally, we note that for separable two-qubit states, concurrence $C$ zero by definition. The local quantum uncertainty (LQU) may also be zero for certain classical-quantum states depending on the measurement basis, while quantum coherence $\mathcal{C}_l$ is zero for states diagonal in the chosen reference basis. Similarly, the Bell–CHSH observable $\mathbb{B}$ cannot violate the inequality ($\mathbb{B} \leq 2$) {for any separable state.} These observations provide a clear benchmark for interpreting the behavior of quantum correlations and coherence in our system, highlighting the conditions under which these measures signal non-classical properties.

Our research demonstrates that manipulating anisotropies, and interaction parameters together enables effective control of quantum resources in spin systems operating at finite temperatures. The three main components of our research, including magnetic and coupling anisotropies and the Dzyaloshinskii–Moriya interaction, provide effective tools for creating and maintaining quantum correlations, and coherence. Maintaining control over quantum systems operating in real-world conditions is critical for developing upcoming quantum technologies, such as solid-state quantum information processing and spintronics.
{From an applied perspective, controlling magnetic and coupling anisotropies from an applied perspective provide an effective method of enhancing the thermal protection of quantum resources against   external disturbances. Tuning these parameters enables the design of quantum devices that exhibit improved performance, and stability. Our  results offer practical guidance for engineering  spin-based quantum systems that maintain coherence, and correlations under realistic experimental conditions. 
Overall, these findings represent an important step toward bringing quantum technologies closer to practical implementation, since they demonstrate that such systems can operate well beyond the idealized zero-temperature limit.}


\end{document}